\documentclass[aps,prx,twocolumn,showpacs,superscriptaddress,longbibliography,10pt]{revtex4-1}

\usepackage[english]{babel}
\usepackage{amsmath,amsfonts,amsthm,braket}
\usepackage{morefloats}
\usepackage{bm}
\usepackage{graphicx}
\usepackage[outdir=./]{epstopdf}

\usepackage{hyperref}
\newcommand*\chem[1]{\ensuremath{\mathrm{#1}}}
\newcommand*\ham{\hat{H}}
\newcommand*\crt[1]{\hat{a}^\dagger_{#1}}
\newcommand*\dst[1]{\hat{a}^{\phantom{\dagger}}_{#1}}
\newcommand*\vett[1]{{\bf{#1}}}
\newcommand*\Ha{\mathrm{E_{Ha}}}
\newcommand*\mHa{\mathrm{m\,E_{Ha}}}
\newcommand*\bohr{\,\mathrm{a_B}}

\newcommand*\tdl{\mathrm{TDL}}
\newcommand*\corr{\mathrm{corr}}

\newcommand{\COMMENTED}[1]{}
\usepackage{lineno}

\newcommand{\REV}[1]
{{#1}}

\newcommand{\REVISED}[1]
{{#1}}

\usepackage[dvipsnames]{xcolor}

\begin{document}

\author{Mario Motta}
\affiliation{Department of Physics, College of William and Mary, Williamsburg, VA 23187, USA}

\author{David M. Ceperley}
\affiliation{Department of Physics, University of Illinois at Urbana-Champaign, Champaign, IL 61801, USA}

\author{Garnet Kin-Lic Chan}
\affiliation{Division of Chemistry and Chemical Engineering, California Institute of
   Technology, Pasadena, CA 91125, USA}

\author{John A. Gomez}
\affiliation{Department of Chemistry, Rice University, Houston, TX 77005, USA}

\author{Emanuel Gull}
\affiliation{Department of Physics, University of Michigan, Ann Arbor, MI 48109, USA}

\author{Sheng Guo}
\affiliation{Division of Chemistry and Chemical Engineering, California Institute of
   Technology, Pasadena, CA 91125, USA}

\author{Carlos Jimenez-Hoyos}
\affiliation{Division of Chemistry and Chemical Engineering, California Institute of
   Technology, Pasadena, CA 91125, USA}

\author{Tran Nguyen Lan}
\affiliation{Department of Chemistry, University of Michigan, Ann Arbor, MI 48109, USA}
\affiliation{Department of Physics, University of Michigan, Ann Arbor, MI 48109, USA}

\affiliation{On leave from: Ho Chi Minh City Institute of Physics, VAST, Ho Chi Minh City, Vietnam}

\author{Jia Li}
\affiliation{Department of Physics, University of Michigan, Ann Arbor, MI 48109, USA}

\author{Fengjie Ma}
\affiliation{Department of Physics, Beijing Normal University, Beijing, Beijing 100875, China}

\author{Andrew J. Millis}
\affiliation{Department of Physics, Columbia University, New York, NY 10027, USA}

\author{Nikolay V. Prokof'ev}
\affiliation{Department of Physics, University of Massachusetts, Amherst, MA 01003, USA}
\affiliation{National Research Center ``Kurchatov Institute'', 123182 Moscow, Russia}

\author{Ushnish Ray}
\affiliation{Division of Chemistry and Chemical Engineering, California Institute of
   Technology, Pasadena, CA 91125, USA}

\author{Gustavo E. Scuseria}
\affiliation{Department of Chemistry, Rice University, Houston, TX 77005, USA}
\affiliation{Department of Physics and Astronomy, Rice University, Houston, TX 77005, USA}

\author{Sandro Sorella}
\affiliation{SISSA -- International School for Advanced Studies, Via Bonomea 265, 34136 Trieste, Italy}
\affiliation{Democritos Simulation Center CNR--IOM Istituto Officina dei Materiali, Via Bonomea 265, 34136 Trieste, Italy}

\author{Edwin M. Stoudenmire}
\affiliation{Department of Physics and Astronomy, University of California, Irvine, CA 92697-4575 USA}

\author{Qiming Sun}
\affiliation{Division of Chemistry and Chemical Engineering, California Institute of
   Technology, Pasadena, CA 91125, USA}

\author{Igor S. Tupitsyn}
\affiliation{Department of Physics, University of Massachusetts, Amherst, MA 01003, USA}
\affiliation{National Research Center ``Kurchatov Institute'', 123182 Moscow, Russia}

\author{Steven R. White}
\affiliation{Department of Physics and Astronomy, University of California, Irvine, CA 92697-4575 USA}

\author{Dominika Zgid}
\affiliation{Department of Chemistry, University of Michigan, Ann Arbor, MI 48109, USA}

\author{Shiwei Zhang}
\altaffiliation{shiwei@wm.edu}
\affiliation{Department of Physics, College of William and Mary, Williamsburg, VA 23187, USA}

\title{Towards the solution of the many-electron problem in real materials: \\                                                                                                 
       equation of state of the hydrogen chain with state-of-the-art many-body methods}       

\collaboration{The Simons Collaboration on the Many-Electron Problem}

\date{\today}
   
\begin{abstract}
We present numerical results for the equation of state of an infinite chain of hydrogen atoms.
A variety of modern many-body methods are employed, with exhaustive cross-checks and validation.
Approaches for reaching the continuous space limit and the thermodynamic limit are investigated, proposed, and tested.
The detailed comparisons provide a benchmark for assessing the current state of the art in many-body 
computation, and for the development of new methods.
The ground-state energy per atom in the linear chain is accurately determined  versus bondlength, with a
confidence bound given on all uncertainties.
\end{abstract}

\maketitle

\section{Introduction}

One of the grand challenges in modern science is the accurate treatment of interacting 
many-electron systems. 
In condensed phase materials, the challenge is increased by the need to account for
the interplay between the electrons and the chemical and structural environment.
Progress in addressing this challenge will be fundamental to the realization of ``materials genome'' or 
materials by design initiatives.
 
 Often the physical properties of materials and molecules are determined by a delicate 
 quantitative balance between competing tendencies, so that accurate computations are required to 
 predict the outcome. 
 The theoretical framework for these calculations, the many-particle Schr\"odinger equation, is known 
 \cite{Dirac1929}.
 However, the solution of the Schr\"odinger equation in a many-electron system presents fundamental 
 difficulties originating from combinatorial growth of the dimension of the Hilbert space involved,
 along with the high degree of entanglement produced by the combination of Fermi statistics and 
 electron-electron interactions.
 Computational methods need to reach beyond the incredible success afforded by density functional 
theory (DFT), and capture electron correlation effects with sufficient accuracy across different physical 
parameter regimes.

No general, numerically exact method presently exists that can treat many-electron systems with low 
 computational cost. Except for special cases,
known methods either have systematic errors which cannot be easily quantified, or
the computational burden scales 
 exponentially or as a very high power of the system size.

Recent years have witnessed remarkable progress in the development of new theories, concepts, methodologies, software 
and algorithms that have pushed the conceptual horizons and technical boundaries of computational many-body methods, 
and considerably improved our understanding of interacting electrons in solids and molecules. 
A vast suite of methods exist which have different strengths and weaknesses and different domains of applicability, 
and ever more are being developed. 

It is imperative, under these circumstances, to develop systematic knowledge by detailed benchmark studies.  
Comparison of different methods allows characterization of relative accuracy and capabilities, which provides a survey of the state-of-the-art to guide applications. 
Applying complementary methods synergistically to the same problem enables cross-check and validation, 
leading to a powerful new paradigm of attack on difficult problems.
Cases where results from different methods agree provide valuable benchmarks against which new 
methods can be tested, thereby facilitating further development and accelerating progress. 

Detailed benchmark studies of extended systems have been rare.
A major reason is the nature of the problems involved: while it is reasonably straightforward to compare 
results obtained for finite systems (such as molecules), or at the thermodynamic limit in an 
independent-electron picture, 
it is challenging to make reliable calculations in the thermodynamic 
limit with many-body methods. A recent success is the benchmark study of the Hubbard model  \cite{LeBlanc2015}, 
and a subsequent multi-method study of the  
ground-state order in the celebrated 
$1/8$-doped case \cite{Zheng2017}. 
With real materials, two more challenges arise. First, the general long-range Coulomb interaction must be treated accurately. 
Second, many-body calculations often require the use of incomplete one-electron basis sets,
whose systematic errors must be removed in order to reach the  
continuous space, or complete basis set, limit for physical observables.

In this work, we undertake a comprehensive benchmark study of state-of-the-art many-body methods in 
a more realistic context. 
We choose the linear hydrogen chain ---
introduced in Ref.~\cite{Hachmann2006}
and studied at finite lengths and finite basis sets by several groups 
\cite{AlSaidi2007,Tsuchimochi2009,Sinitski2010,Mazziotti2011,Lin2011,Sorella2011}
--- and investigate its ground state versus bondlength under the Born-Oppenheimer approximation
of fixed nuclear positions, at the thermodynamic and complete basis set limits.

Hydrogen is the first element in the periodic table
and the most abundant in nature. Studies of the H atom, H$_2^+$ cation and H$_2$ molecule have served as
landmarks in quantum physics and chemistry. Despite their deceptive simplicity, bulk H systems are
rich and complex. 
The ground state properties of the hydrogen chain can differ significantly from those of simpler systems
such as the Hubbard model, and are, in fact, not completely understood. The linear H chain captures key
features that are essential to the generalization of model-system methods to real materials,  
in particular strong electron correlations of diverse nature arising as the H-H distance is increased, 
the need to treat the full physical Coulomb interaction, and to work in the continuous space and thermodynamic limits.

Compared to the one-dimensional Hubbard model, the
hydrogen chain has multiple (in principle, infinite) orbitals per site, as well as long-range interactions.
The use of basis sets defines models of the hydrogen chain of increasing complexity. In a minimal basis, there is
only one band, and the problem resembles a one-dimensional Hubbard model with long-range interactions.
Larger, more realistic, basis sets bring back characteristics of real materials. Thus one can neatly and systematically 
connect from a fundamental model of strong electron correlation to a real material system.
On the other hand, the H chain eliminates complexities of other materials systems such as the need to separately treat core 
electrons or incorporate relativistic effects, and is thus accessible to many theoretical methods at their current state of development. 
As such, the linear hydrogen chain is an ideal 
first benchmark system for testing the ability of many-body theoretical methods to handle the challenges posed by real materials. 

We study finite chains of increasing length, and cross-check the results against calculations performed using 
periodic boundary conditions. 
We also present results from calculations formulated 
in  the thermodynamic limit.
Most of the methods employed use a one-particle basis set, and we investigate convergence by obtaining results 
for a systematic quantum chemistry sequence of basis sets of increasing size.
These results are compared to methods formulated directly in real space. 
With extensive direct comparisons and cross-validations between different methods, we characterize the 
uncertainties in each approach in detail.

This study presents results obtained from more than a dozen many-body computational methods currently used 
or under development in physics and chemistry.
A vast amount of data is produced, providing detailed information in finite-length chains and with finite basis sets.
In the largest systems treated, the size of the Hilbert space exceeds $10^{100}$.
We anticipate that our data, which are made available in the appendices 
and in online repositories \cite{github2017},
 will be useful for benchmarking other existing and future electronic structure methods.
 \REVISED{In addition to the results and comparisons, we introduce a variety of methodological
 developments which were spurred by the benchmark, including new approaches.}
Combining the strengths of complementary methods, we are able to determine the 
energy per atom 
in the thermodynamic limit to sub-milli-Hartree  accuracy. 
We hope that the results presented here will serve as a 
preview of what can be achieved in the predictive computation of the properties
of real materials, and 
provide a firm basis for theoretical progress in condensed matter physics, quantum chemistry, materials science and related fields.

The rest of the paper is organized as follows.
In Sec.~\ref{sec:model} we introduce the linear \chem{H} chain systems that will be studied and define notation. 
In Sec.~\ref{sec:methods} we give a brief overview of the many-body methods employed in the present work. 
More details on the methods and some of the computational details of each method are given in 
Appendix~\ref{sec:methods-description}.
In Sec.~\ref{sec:h10} we present results for a finite \chem{H_{10}} chain. 
A subsection summarizes the results within a minimal basis set, followed by one which presents the results in the complete 
basis set (CBS) limit, and by one that describes the extrapolation to the CBS limit from finite basis set results.
In Sec.~\ref{sec:tdl} we present results 
in the thermodynamic limit.
The first two subsections contain results for the minimal basis and the CBS limit, respectively, while the last subsection 
presents finite-size results and discusses the procedure for reaching the thermodynamic limit.
A brief discussion and summary of our work, along with future prospects, is then given in Sec.~\ref{sec:conclusion}.
In the Appendices we include further descriptions of the methods, present tables that summarize our data, and provide further details of our benchmark 
results and procedures. 
A database of results is also made available electronically \cite{github2017}.

\section{The Hydrogen chain}
\label{sec:model}

We consider a system comprised of $N$ protons, at fixed equispaced positions along a line, with $N$ electrons.
This system is described by the Hamiltonian
\begin{equation}
\label{eq:ham1st}
\begin{split}
\ham = &-\frac{1}{2} \, \sum_{i=1}^N \nabla^2_i + \sum_{i<j=1}^N \frac{1}{|{\bf{r}}_i - {\bf{r}}_j|} \\
            &- \sum_{i,a=1}^N \frac{1}{|{\bf{r}}_i - {\bf{R}}_a|} + \sum_{a<b=1}^N \frac{1}{|{\bf{R}}_a - {\bf{R}}_b|} \,,\\
\end{split}
\end{equation}
where $(\vett{r}_1 \dots \vett{r}_N)$ and $(\vett{R}_1 \dots \vett{R}_N)$ are the coordinates of electrons and nuclei, respectively.
We will use atomic units throughout, i.e., lengths are measured in Bohr ($\bohr =\hbar^2/(m_e e^2)$) and energies
in Hartree ($\Ha=e^2/a_B$).
In the thermodynamic limit of infinite 
system size at zero temperature, 
which is our primary focus, such a system is characterized by only one parameter, 
the bondlength $R$ separating two adjacent atoms.

The electron coordinates are continuous in three-dimensional space, while the nuclear coordinates are 
fixed on a line, e.g., ${\bf{R}}_a =a\,R\,{\bf{e}}_z$ with $R$ the inter-proton separation and $a=1...N$.
Most of our calculations are on such finite-size systems, referred to as open boundary conditions (OBC). 
We have also performed calculations using periodic boundary conditions (PBC), in which case a periodic supercell 
containing $N$ atoms is treated.

Among the methods employed here, diffusion Monte Carlo (DMC) and variational Monte Carlo (VMC) operate 
in first-quantization and  treat the Hamiltonian in Eq.~(\ref{eq:ham1st}) directly. The other methods use a finite 
one-electron basis set, with a total of $M$  orbitals $\{ \varphi_p \}_{p=1}^M$, 
i.e., $m\equiv M/N$ basis functions  \emph{per atom} 
(including occupied and virtual orbitals).
The Hamiltonian is written in second-quantized form  
\begin{equation}
\label{eq:ham2nd}
\ham = \sum_{pq=1}^M h_{pq} \crt{p} \dst{q} + \frac{1}{2} \sum_{pqrs=1}^M v_{pqrs} \, \crt{p} \crt{q} \dst{r} \dst{s}\,,
\end{equation}
where the creation and annihilation operators 
$\crt{}$ and $\dst{}$ 
obey fermionic anticommutation relations and the indices $p$,$q$,$r$,$s$ run over all $M$ single-electron basis functions. 

Most calculations were performed using standard Gaussian basis sets, but
 specialized density-matrix renormalization group (DMRG) calculations using a grid along the nuclear axis and Gaussians along the
other two directions (sliced-basis) were also performed, as discussed in Sec.~\ref{sec:dmrg}.
 Within the Gaussian basis, the matrix elements $\{h,v\}$ in Eq.~(\ref{eq:ham2nd}) are readily obtained from standard 
 quantum chemistry packages. 
 The basis functions are centered on the protons.
 We use the correlation-consistent cc-pV$x$Z basis set, with $x=$D,T,Q, and 5, 
which correspond to 
$m=$5, 14, 30, and 55 
orbitals 
per atom,
 respectively.
 For small $N$, explicit correlation using the F12 technique \cite{Kutzelnigg1985,Kong2010} was also considered 
to help ascertain the approach to the CBS limit. 
Our procedure for extrapolating to the CBS limit is described in 
Secs.~\ref{subsec:h10-extrap-x} and \ref{sec:tdl-extrp-x-N}.

 H-chains with nearest-neighbor proton separation (bondlength) $R$ of 1.0,   1.2, 1.4, 1.6, 1.8,  2.0, 2.4, 2.8, 3.2, and 3.6\,Bohr are studied.
We focus, in this work, on the ground-state energy $\mathcal{E}(N,R)$  for different chain sizes and lengths, and obtain
the \emph{energy per atom}, $E(N,R) = \mathcal{E}(N,R)/N$, at the thermodynamic limit  (TDL)
\begin{equation}
E_{\tdl}(R) = \lim_{N \to \infty} E(N,R) \, .
\end{equation}
Below when presenting results on finite chains, we will follow the chemistry convention and use
the term potential energy curve (PEC), although we will always refer to energy per atom, $E(N,R)$.
When presenting results for the TDL, we will use the term equation of state (EOS) to refer to 
$E_{\tdl}(R)$ vs.~$R$ at zero temperature.
Most of the methods considered
chains with $N=10$-$102$ atoms.
The procedure for extrapolating to  
 $N\to \infty$ is discussed in Sec.~\ref{sec:tdl-extrp-x-N}.

\section{Overview of computational Methods}
\label{sec:methods}

\begin{table*}[t!]
\begin{tabular}{lllllll}
\hline\hline
\multicolumn{2}{c}{method} & deterministic & basis set & self-consistent & variational & scaling                                      \\
\hline
Wave-function 
&CCSD                      & yes           & b         & yes             & no          & $N^2 M^4 + N^3 M^3$                          \\
&CCSD(T)                   & yes           & b         & yes             & no          & $N^3 M^4$                                    \\
&DMRG                      & yes           & b         & yes             & yes         & $D^3 M^3 + D^2 M^4$                          \\
&SBDMRG                    & yes           & sb        & yes             & yes         & $N R D^3 \left[ N_o^3 + D(N_o) \right]$      \\
&HF                        & yes           & b         & yes             & yes         & $M^4$                                        \\
&FCI                       & yes           & b         & no              & yes         & $\binom{M}{N}$                               \\
&MRCI                      & yes           & b         & no              & yes         & $> \binom{N}{N/2} N^4 + N^2 M^4$             \\
&NEVPT2                    & yes           & b         & no              & no          & $\binom{N}{N/2} N^8$                         \\
\cline{2-7}                                                                                                                           
&AFQMC                     & no            & b         & no              & no          & $N^2 M^2 + M^2 N$                            \\
&VMC                       & no            & cs        & no              & yes         & $N^2 M + N^3$                                \\
&LR-DMC                    & no            & cs        & no              & yes         & $N^2 M + N^3$                                \\
\hline                     
Embedding                  
&DMET                      & yes           & b         & yes             & no          & $N_f^3 D^3 + N_f^2 D^4 \left [(N_f^3 D^3 + N_f^2 D^4) M\right]$              \\
&SEET                      & yes           & b         & yes             & no          & $N_{imp}\binom{M_s}{n_e} +  M^5 n_\tau  \  [N_{imp}\binom{M_s}{n_e} +  M^4]$ \\
\hline                     
Diagrammatic               
&SC-GW                     & yes/no        & b         & yes             & no          & $M^4 n_\tau$                                 \\
&GF2                       & yes           & b         & yes             & no          & $M^5 n_\tau$                                 \\
&BDMC$_n$                  & no            & b         & yes             & no          & $e^{n\ln n}$                               \\
\hline\hline
\end{tabular}
\caption{Summary of the methods employed in the present work. 
Methods are classified by type: wavefunction,  embedding, or diagrammatic; 
the use of one-electron Gaussian basis sets (b), sliced-basis sets (sb), 
or continuous electron positions (cs);
whether a self-consistency procedure is involved; 
whether the method is deterministic or stochastic in nature; and
whether or not the method is variational.
The scaling of the computational cost of key pieces of the algorithm is shown, 
versus 
the number of electrons ($N$), basis set size ($M$), etc. 
In DMRG, $D$ is the bond dimension kept in the calculation. 
In SBDMRG, $N_o$ is the number of basis function per slice and $D(N_o)$ 
the compressed MPO size.
In Green's function methods, $n_\tau$ is the size of the imaginary time grid.
In DMET, 
$N_f$ is the number of atoms in the fragment and $D$ denotes the bond dimension kept by the DMRG solver.
The first scaling corresponds to a DMET calculation with a single fragment using translational invariance, while the second corresponds
to treating multiple fragments tiling the chain.
In SEET, $N_{imp}$ is the number of impurities, while $n_e$ denotes the number of electrons in the impurity
and $M_s=M^A+M_b$ the number of orbitals treated, where $M_b$ is the number of bath orbitals. The first scaling corresponds to SEET(CI/GF2), while the second corresponds to SEET(CI/HF).
In BDMC, $n$ denotes the diagrammatic order. 
}
\label{tab:classification}
\end{table*}

The methods employed in this work include: 

\begin{itemize}

\item AFQMC: auxiliary-field quantum Monte Carlo \cite{Blankenbecler1981,Zhang1997,Zhang2003}  

\item BDMC: bold diagrammatic Monte Carlo  \cite{VanHoucke2010,Kulagin2013,Phonons2016} 

\item DMET: density matrix embedding theory \cite{Knizia-2012,Knizia-2013,Wouters2016} 

\item DMRG: density matrix renormalization group 
with a quantum chemistry basis \cite{white1992density,white1999ab,chan2002highly,chan2016matrix}

\item FCI: full configuration interaction, i.e., exact diagonalization

\item GF2: self-consistent second-order Green's function \cite{Phillips14,rusakov_gf2_periodic,fractional,GF2_thermo,dahlen04,dahlenmp2}

\item LR-DMC: lattice-regularized \cite{lrdmc} diffusion Monte Carlo (DMC) \cite{reynolds1982}

\item MRCI, MRCI+Q: multireference 
configuration interaction without (MRCI) and with (MRCI+Q) the Davidson correction \cite{werner1988efficient,knowles1988efficient}

\item NEVPT2: 
partially (PC-) and strongly contracted (SC-) variants of the N-electron valence state second order perturbation 
theory \cite{angeli2007new} 

\item RCCSD and RCCSD(T): coupled cluster (CC) theory with full treatment of singles and doubles
(RCCSD ) and perturbative treatment of triple excitations (RCCSD(T)), using RHF as a reference state \cite{Paldus1999,Bartlett2007,Shavitt2009}

\item RHF: restricted Hartree-Fock \cite{Roothaan1960}

\item SBDMRG: specialized DMRG with sliced basis sets \cite{Stoudenmire:2017}

\item SC-GW: fully self-consistent GW \cite{Almbladh1999,Stan2006,stan09,Phillips14}

\item SEET: self-energy embedding theory  \cite{AlexeiSEET,LAN,nguyen2016rigorous,Zgid_Gull_seet_general,lan_mixing}

\item UCCSD and UCCSD(T): CC theory with full treatment of singles and doubles
(UCCSD ) and perturbative treatment of triple excitations (UCCSD(T)), using UHF as a reference state \cite{Paldus1999,Bartlett2007,Shavitt2009}

\item UGF2: spin-unrestricted GF2 \cite{fractional}

\item UHF: unrestricted Hartree-Fock \cite{Pople1954}

\item VMC: variational Monte Carlo \cite{McMillan1965,Ceperley1977}

\end{itemize}

Detailed descriptions of the
methods and specific calculational details are presented in Appendix~\ref{sec:methods-description}.
We focus on many-body methods in this work. The independent-electron methods  RHF and UHF are listed above,
because they are used by many of our methods as reference, initial, or trial states. 
Further, we will use
the RHF 
to define the correlation energy for the purpose of extrapolation to the CBS limit
[regardless of the nature of the 
mean-field reference (if any) used in the method].

In Table~\ref{tab:classification} we summarize the methods using several 
characteristics or criteria.
At a high level, the methods can be distinguished by general categories such as 
wave-function, embedding, and diagrammatic. 
Wave-function methods (AFQMC, CC, DMRG, FCI, LR-DMC, MRCI, NEVPT2, and  VMC) formulate an ansatz for the ground state, and 
compute expectation values of observables and correlation functions with respect to the wave function.
The ansatz for the wave function can be explicit (as in VMC and most quantum chemistry methods), or 
reached 
via an iterative procedure (as in AFQMC, LR-DMC). 
The accuracy of a wave-function method is determined by the quality of the underlying ansatz 
(e.g., form of trial wave function in VMC, size of truncated space, order of perturbation)
and by approximations (if any) in the realization of the ansatz 
(e.g., constraints in QMC) and in the 
evaluation of observables (e.g., non-variational estimators in CC methods, mixed estimate or back-propagation in QMC).
Extrapolations in $N$ (and, in many cases, basis set size $M$) are needed for wave-function methods.

Embedding methods (DMET, SEET) evaluate the properties of a large system by partitioning it within a basis, e.g. the spatial or energy basis,
into a collection of fragments, embedded in a self-consistently determined environment treated at a more approximate level.
The accuracy of an embedding method is determined by a combination of  several factors including: the size of the fragments,
the level of accuracy in the treatment of the embedded fragments and environment, and the level of convergence of the self-consistency loop.
Extrapolations in the fragment size (and in the basis set size $M$ if a basis set is used) is needed for embedding methods. 

Diagrammatic methods (BDMC, GF2, SC-GW) evaluate, either deterministically (GF2, SC-GW) or stochastically (BDMC), terms in a diagrammatic expansion of a  system property. 
Diagrammatic expansions can be formulated either in a basis or directly in the continuum, and can be applied to finite systems or directly in the thermodynamic limit. GF2 provides an exact self-consistent evaluation of all second order 
terms in a perturbative expansion in the interaction; SC-GW evaluates a small subset of diagrams to all orders in perturbation theory via the solution of self-consistent equations; 
 BDMC accounts for higher-order vertex corrections within the skeleton expansion by performing a stochastic sampling of diagram topologies and internal integrations, and is limited to situations where the series converges.
(We note an important ambiguity in the formulation of diagrammatic approximations: 
Hamiltonian terms that are identically zero because of the Pauli exclusion principle give rise to diagrams that 
may not sum to zero (self-interaction error) in approximations that do not consider all terms at a given order. 
The effect of these zero-terms on SC-GW results is illustrated in Appendix~\ref{sec:diagrammatic}.)

Many other types of classification are possible.
For example, one could characterize a method as deterministic or stochastic; 
whether a basis set is used and if so, what type; whether a 
self-consistent procedure is involved; whether the computed ground-state energy is variational; etc.
Table~\ref{tab:classification} lists some of these classifiers, in addition to the computational scaling of key pieces
in each algorithm. It is important to note that the classification is only meant as a general guide, and is 
in many cases fluid. For instance, embedding methods could also be classified by their solver for the 
embedding fragments, as wave function (DMET) or diagrammatic (SEET). Depending on the particular form of the solver, 
they could be deterministic (e.g., DMET with a DMRG solver as in the present study) or stochastic 
(if a QMC solver is used \cite{DMET-AFQMC}). Various methods are shown as needing a Gaussian 
basis set, but can also be implemented using other bases (e.g., plane-wave with pseudopotentials
for AFQMC, and SC-GW). The choice of basis set can affect the computational scaling.
Note also that the meaning of self-consistency can vary and depend on the type of methods.  
In wave-function methods, we have used it to indicate whether a self-consistent procedure is involved, 
although this can still have ambiguity since there are sometimes multiple ways to obtain the solution.
Finally, the scaling reported  refers to the canonical implementation of these methods, without any 
specialized optimizations.

\section{Results for the \chem{H_{10}} chain}
\label{sec:h10}

In this section, we present results on a finite chain of ten atoms.
This relatively simple system provides a good intermediate step in the benchmark, as it removes one of 
the major challenges, namely the approach to the thermodynamic limit. Detailed comparisons can be
made as in quantum chemistry, providing insights for the more challenging case of the TDL. We 
emphasize the approach to the continuous space limit, with extensive studies on the removal of any residual 
finite basis errors. In Sec.~\ref{subsec:h10-min},
results in the minimal basis are given, for which exact results from FCI are available for detailed comparison.
Final results are presented for the CBS limit in Sec.~\ref{subsec:h10-CBS}. Then, in Sec.~\ref{subsec:h10-extrap-x}, we include results using finite basis sets and discuss the approach to the CBS limit. 

\subsection{Benchmark in the minimal basis set}
\label{subsec:h10-min}

\begin{figure}[h!]
\centering
\includegraphics[width=0.9\columnwidth]{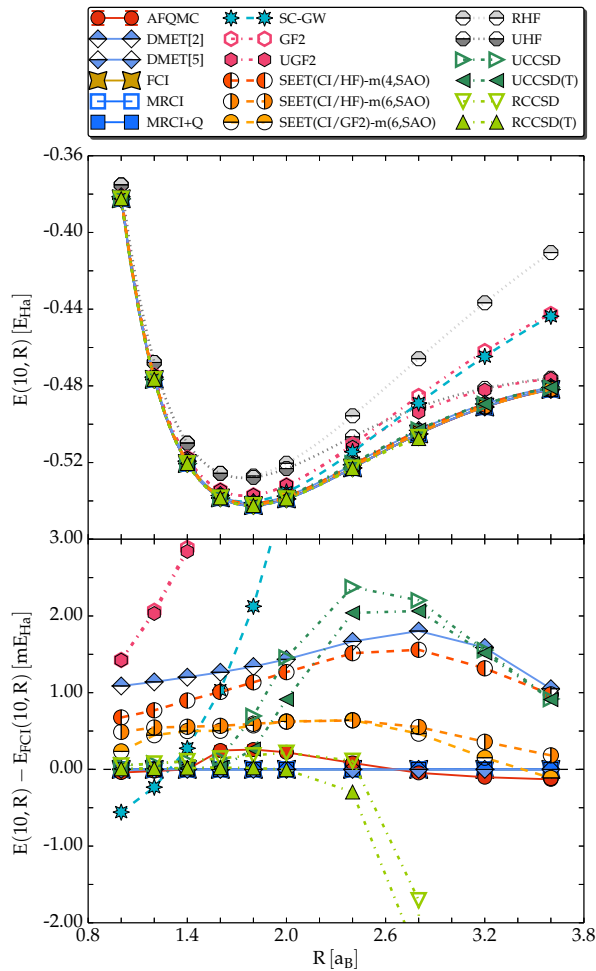}
\caption{
Potential energy curve of \chem{H_{10}} (top) and deviations from FCI (bottom), in the minimal STO-6G basis.
}\label{fig:H10_sto6g}
\end{figure}

Figure~\ref{fig:H10_sto6g} shows a detailed comparison of the potential energy curve (PEC),
$E(N=10,R)$ vs.~$R$, 
obtained by 
a variety of methods,
in the minimal STO-6G basis.
For all methods, the PEC 
features a familiar short-range repulsion, due to the combined effect of Coulomb repulsion
and Pauli exclusion, followed by a decrease to a minimum value $E_0$, attained at the equilibrium bondlength $R_e$.
Beyond $R_e$, the PEC monotonically increases to an asymptotic value $E_\infty$, the 
ground-state energy of a single \chem{H} atom. 
The well depth gives a dissociation energy $D_e = (E_\infty - E_0)$.
Owing to the small size of this chain and the STO-6G basis, the PEC can be calculated using the FCI method, 
giving the exact values $R_e = 1.786\,\bohr$, $E_0 = -0.542457\,\Ha$, and $E_\infty=-0.471039\,\Ha$.

The overall agreement between all the many-body methods is quite good. 
Deviations from FCI are shown in the lower panel of Fig.~\ref{fig:H10_sto6g}
 in a magnified view.
We see that the different methods agree with each other and with the exact result at the
level of $\sim$ 2 $\mHa$, i.e. about $5 \%$ of the dissociation or cohesive energy. 
The agreement is better for $R \leq 2.0 \bohr$ and tends to worsen in many methods as $R$ increases, because electronic 
correlations become more pronounced, increasing the multireference character of the system.

MRCI and especially MRCI+Q are seen to be uniformly accurate for this system, with discrepancies 
of $\mu\,\Ha$ or less from FCI. This is also confirmed 
 in the larger cc-pVDZ basis set (see Sec.~\ref{subsec:h10-extrap-x}), where 
MRCI+Q and DMRG results are virtually indistinguishable. 
 MRCI+Q can be conveniently carried out for even larger basis sets (but not for large $N$), and we will use it 
as our reference in the following subsections for the 10-atom chain.

Among the other approaches, AFQMC gives results accurate to within $0.2\,\mHa$. Bias from the CP approximation is visible in the intermediate region,
where the energy is slightly overestimated, and also at large bondlengths, where there is an underestimation. 
Coupled cluster methods, especially RCCSD(T), are very accurate near equilibrium. 
Although it is in principle possible to dissociate
the H chain to the correct energy within RCCSD (as a product of dissociated H dimers) there can be multiple CC solutions, and in practice 
a correctly dissociating solution is hard to find \cite{Paldus1999}. UCCSD and UCCSD(T) provide accurate results at equilibrium and approach
the correct result at longer bond lengths, but have large errors at the intermediate bond-lengths due to spin recoupling.

In embedding methods, extending the number X of embedded atoms (in DMET[X]) or the number Y of impurity orbitals (in SEET-m[Y]) leads to 
noticeable improvement. In the minimal basis, DMET[X] uses $X$ impurity orbitals because there is only one orbital per atom.
The maximum error in DMET[2] is about $2\,\mHa$ while DMET[5] is exact by construction.
While the SEET-m[4] curve is at a similar level of accuracy to DMET[2],  a substantial improvement is obtained within the mixing scheme (SEET-m[6] curves),
especially at large bondlengths. 

As weak coupling methods, the diagrammatic GF2 and SC-GW methods have difficulties in the strong coupling regime at large bondlengths. 
Allowing methods to break spin symmetry may lead to an improvement of the energetics. As illustrated with GF2, using an unrestricted reference
state provides a better estimate of the ground state energy in that regime but generates a spurious magnetization.
Deviations at small distances (corresponding to the weak coupling regime) show that terms beyond the bare second order or screened 
first order approximation are needed to reach the  accuracy of other methods. We also note that the cancellation of self-interaction error in SC-GW 
is subtle and depends on the treatment of exclusion principle violating terms in the Hamiltonian (see Appendix \ref{app:zeroorder}).

\subsection{Potential energy curve in the complete basis set limit}
\label{subsec:h10-CBS}

\begin{figure}[t!]
\centering
\includegraphics[width=0.9\columnwidth]{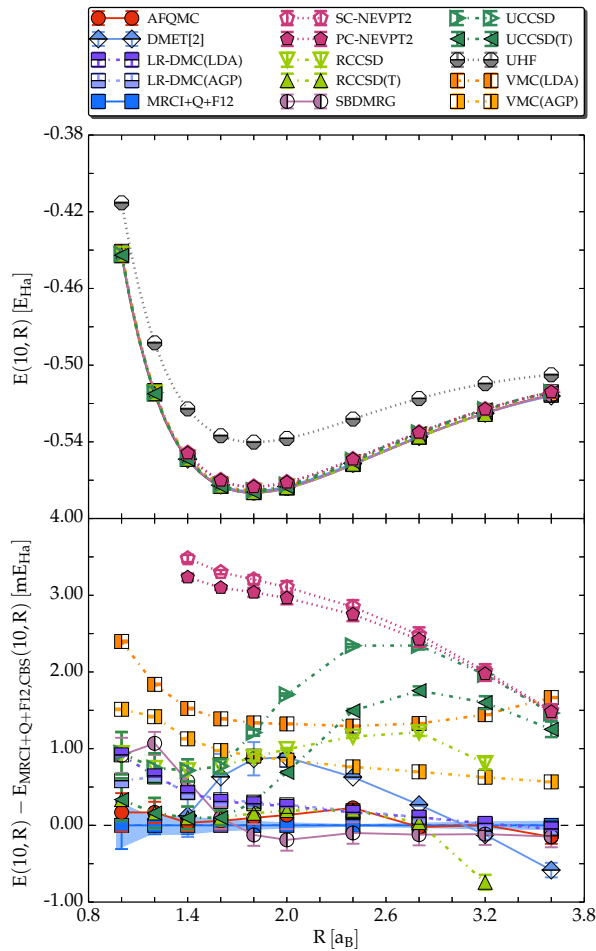}
\caption{
Top: PEC of \chem{H_{10}} in the 
continuous space, or complete basis set (CBS), limit.
Bottom:
Detailed comparison of the final \chem{H_{10}} PECs, 
using MRCI+Q+F12 results at CBS as reference. 
} \label{fig:h10_cbs}
\end{figure}

In Figure~\ref{fig:h10_cbs}, we show the final computed potential energy curves 
of \chem{H_{10}} in the continuous space (complete basis set) limit, including 
results obtained from VMC and LR-DMC, which work in continuous space.

For all our methods that require a basis set, we employ
the correlation-consistent polarized valence $x$-zeta (cc-pV$x$Z) sequence \cite{Dunning1989}, 
which is designed to include successively larger shells of polarization functions
($x=2,3,4,5$ corresponding to D,T,Q,5 respectively).
The results are extrapolated to the CBS limit, following procedures described in Sec.~\ref{subsec:h10-extrap-x}
with further details given in the Appendix. 

Our final results in this system give
an equilibrium geometry  $R_{e} = 1.801(1)$ and energy $E_{0} = -0.5665(1)$.
 The computed PECs are tabulated in the Appendix. 
Deviations from the reference curve are shown in the bottom panel, where the combined uncertainties in the
the reference curve (primarily from the extrapolation to the CBS limit) are indicated. 
Our reference curve for this system was obtained from 
MRCI+Q, extrapolated to the CBS with basis sets up to $x=5$. This is confirmed by a separate extrapolation including 
an   
F12 correction, which gave results in agreement within the statistical uncertainties.

Trends similar to the minimal basis results are observed. 
LR-DMC, which works in continuous electron coordinate space,
has only a weak dependence on basis sets originating from the representation of the Slater determinant in 
the trial wave function (and hence the position of the nodes).
LR-DMC provides an upper bound for the ground-state energy. Its
 quality is determined by the nodal surface of the trial wavefunction.
 At large bondlength, the nodal structure is simpler, consistent with the more quasi-one-dimensional nature
 of the system. The LR-DMC results are  very accurate in this regime, indicating that  the DFT-LDA determinant
 gives a good description of the nodal structure. The AGP trial wave function
 allows a more sophisticated, multi-determinant description of the many-body nodes. 
Improvement with the AGP trial wave function is only seen at the smallest bondlength. 
 The excellent consistency between the LR-DMC results and the basis-set methods provides another 
 assurance on the robustness of the approach to the CBS limit in the latter. 
 
\begin{figure*}[t!]
\centering
\includegraphics[width=0.75\textwidth]{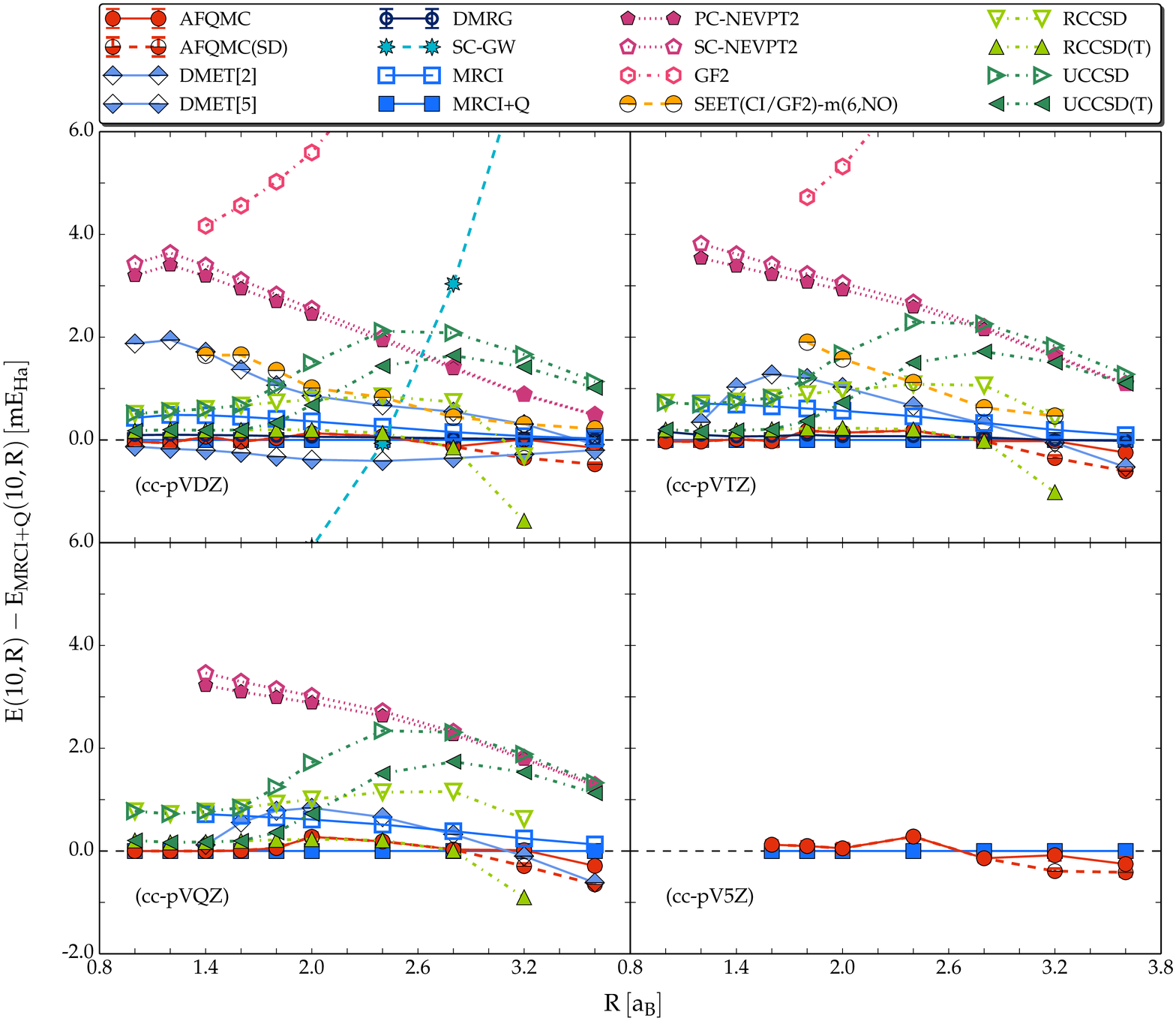}
\caption{
Detailed comparison of the \chem{H_{10}} PEC in each basis set (cc-pVxZ, x=2,3,4,5), using MRCI+Q as reference.
} \label{fig:H10_xZ}
\end{figure*}

\subsection{Reaching the complete basis set limit}
\label{subsec:h10-extrap-x}

For each method which utilizes a basis set, the computational cost grows as the basis size $M$ is increased,
in some cases very rapidly. 
In Fig.~\ref{fig:H10_xZ} we show the PEC yielded by several methods, in the cc-pV$x$Z bases, 
taking MRCI+Q as reference. 
The accuracy of MRCI+Q  is further validated by its excellent agreement with DMRG.
The general trends seen at the minimal basis level are mostly confirmed with the larger basis sets. 
Most methods show errors that remain consistent throughout this family of basis sets which, though
not surprising, is reassuring. 
As mentioned, improvements of the results are possible within certain methods, via larger embedding clusters,
 using better trial wave functions, or going to higher orders. Examples are shown for DMET and SEET;
 for instance, at the cc-pVDZ level, 
 increasing the number of embedded atoms from DMET[2] to DMET[5] reduces the maximum error from $\sim 2 \mHa$ to 
 $0.5 \mHa$. (In the cc-pVTZ and cc-pVQZ bases, cancellation of errors
 means  that the maximum DMET[2] error is $\sim 1 \mHa$.)
Since SEET(CI/GF2) works in the energy basis, increasing the number of impurity orbitals results in significant improvements.
The mixing scheme, which illustrates the strong correlation present in the active space, recovers results of NEVPT2 
quality while solving impurity problems with only 6 orbitals.

In AFQMC, a multideterminant trial wavefunction is used in the dissociation region (last two points, $R>3$) as discussed in Sec.~\ref{sec:methods-description},
which improved its accuracy by $\sim$0.36\,$\mHa$ 
over that with the UHF trial wave function (shown as half-filled red circles).
For CC methods, the improvement from the inclusion of the perturbative triples is systematic and evident.
MRCI+Q energies for the shortest bondlengths and large basis sets relied on a correction as discussed 
in Appendix \ref{sec:numeric}.

\begin{figure}[h!]
\centering
\includegraphics[width=0.99\columnwidth]{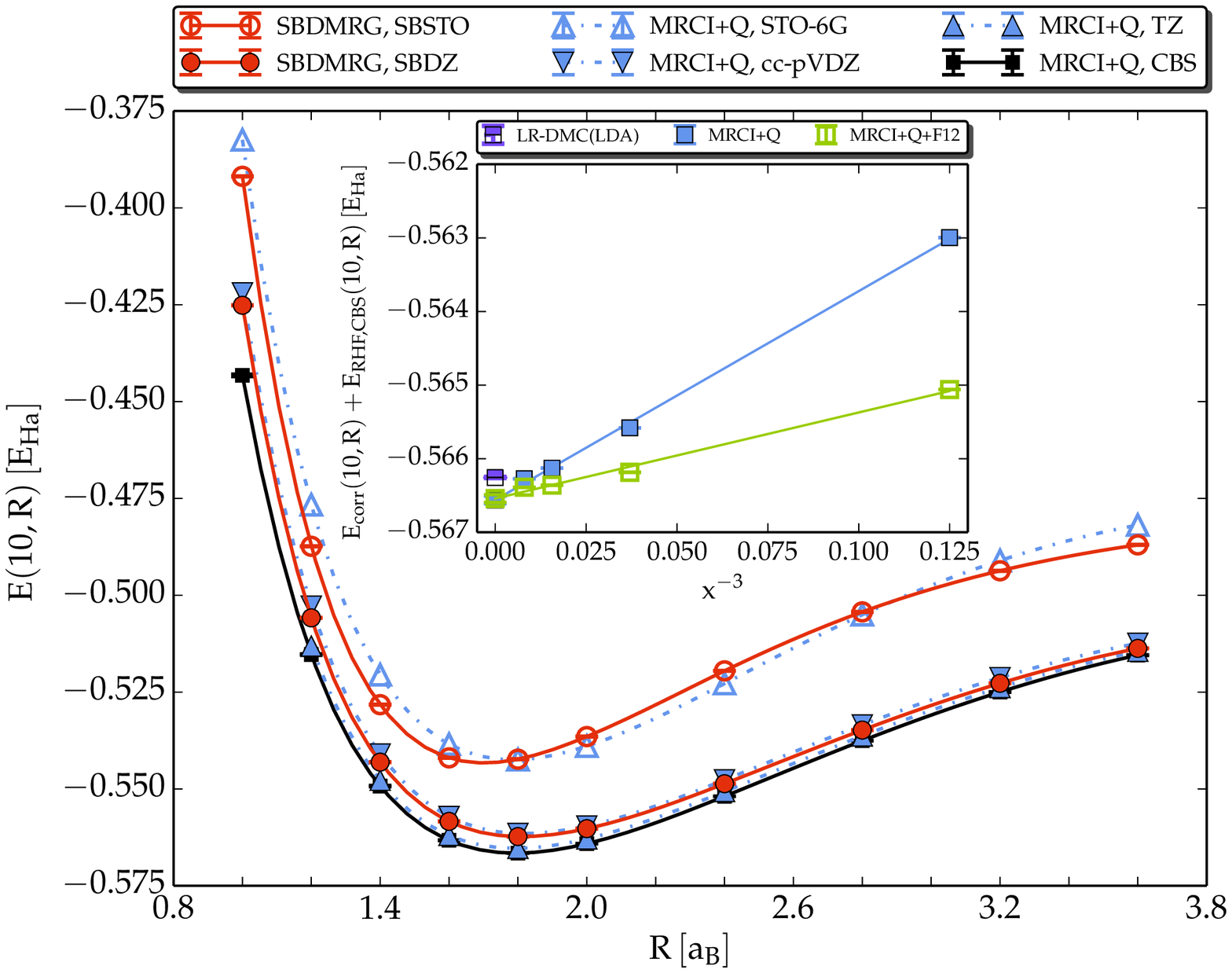}
\caption{
Basis set dependence of the PEC in \chem{H_{10}} and extrapolation to the CBS limit.
MRCI+Q and SBDMRG results 
are shown in the main figure for selected basis sets (to avoid cluttering).
The inset shows the extrapolation of the correlation energy (from MRCI+Q) for 
two sequences of basis sets, cc-pVxZ and with F12, together with LR-DMC, for $R=1.8$.
} \label{fig:H10_mrciq}
\end{figure}

Increasing the basis set size has a dramatic effect on the total energy, as seen in Fig.~\ref{fig:H10_mrciq}.
The basis set dependence is stronger at short bondlengths, with an energy difference  of  $\sim21\,\mHa$ between 
cc-pVDZ and CBS, compared to  $\sim3\,\mHa$ in the bond breaking regime.

The effect of the sliced basis used in the SBDMRG method is also illustrated in the main panel in
Fig.~\ref{fig:H10_mrciq}.  At the STO-6G level there are several competing effects which account for the deviations between the sliced basis and standard basis results.  In the large $R$ limit, the single basis function of STO-6G poorly describes an isolated hydrogen atom, and the increased flexibility in the chain direction of the sliced basis can partially compensate for this. At shorter distances, the overlapping basis functions between adjacent atoms of standard STO-6G give additional degrees of freedom in the transverse directions, which can improve the energy at both the Hartree-Fock level and in terms of transverse correlations. In contrast, the sliced basis set has nearly ideal resolution in the longitudinal direction.
At very short distances, the STO-6G basis becomes nearly linearly 
dependent while the   sliced basis does not and consequently performs significantly better.
These complicated competing transverse and longitudinal effects make it unsurprising that the differences between the two energies changes sign as a function of $R$.
For the cc-pVDZ bases, the sliced version is uniformly slightly better, probably because the dominant effect is its improved longitudinal correlation.

We extrapolate the finite basis set results to the CBS limit by standard procedures \cite{Feller1992,Helgaker1997}, 
taking care to reach large basis sets. 
We first fit the RHF energies $E_{\mathrm{RHF},x}(N,R)$ computed at the cc-pV$x$Z basis set level, 
to an exponential function
\begin{equation}
E_{\mathrm{RHF},x}(N,R) = A(N,R) + B(N,R) e^{ - x C(N,R) }\,.
\end{equation}
The correlation energy
\begin{equation}
E_{\corr,x}(N,R)\equiv E(N,R)-E_{\mathrm{RHF},x}(N,R)
\end{equation}
is then fitted to a power law:
\begin{equation}
\label{eq:fit_corr}
E_{\corr,x}(N,R) = \alpha(N,R) + \frac{\beta(N,R)}{x^3}\,.
\end{equation}
The CBS result is taken as $\alpha(N,R)+A(N,R)$, with a combined uncertainty estimated from the fitting procedures.
We find that using UHF as reference gives numerically indistinguishable results, except for very short bondlengths and large sizes, where convergence of UHF shows more sensitivity.

To deal with basis set linear dependence, which becomes relevant at the shortest bondlengths and largest basis sets, 
we apply a threshold to eigenvalues of the overlap matrix. Threshold values are given under method descriptions in the Appendix.

The final CBS results are verified with a separate set of MRCI+Q calculations augmented by F12 explicit correlation, as illustrated in the 
inset in Fig.~\ref{fig:H10_mrciq}. This data is extrapolated using the same procedure as above.
That these results are consistent with those from the continuous-space LR-DMC provides a  further validation 
of the procedure.

\section{Results in the thermodynamic limit}
\label{sec:tdl}

In this section we present our results in the TDL.
The section is structured similarly to the previous one.
Results for the minimal basis, which makes the H chain resemble an extended Hubbard model, can be valuable for model studies and are described in some detail 
 in Sec.~\ref{sec:tdl-min}.  
The final results for the hydrogen chain at the joint continuous space and thermodynamic limits are given in Sec.~\ref{sec:cbstdl}.
Sec.~\ref{sec:tdl-extrp-x-N} discusses our procedures and cross-validations for approaching 
the thermodynamic limit.

\subsection{Benchmark in the minimal basis set}
\label{sec:tdl-min}

\begin{figure}[h!]
\centering
\includegraphics[width=0.99\columnwidth]{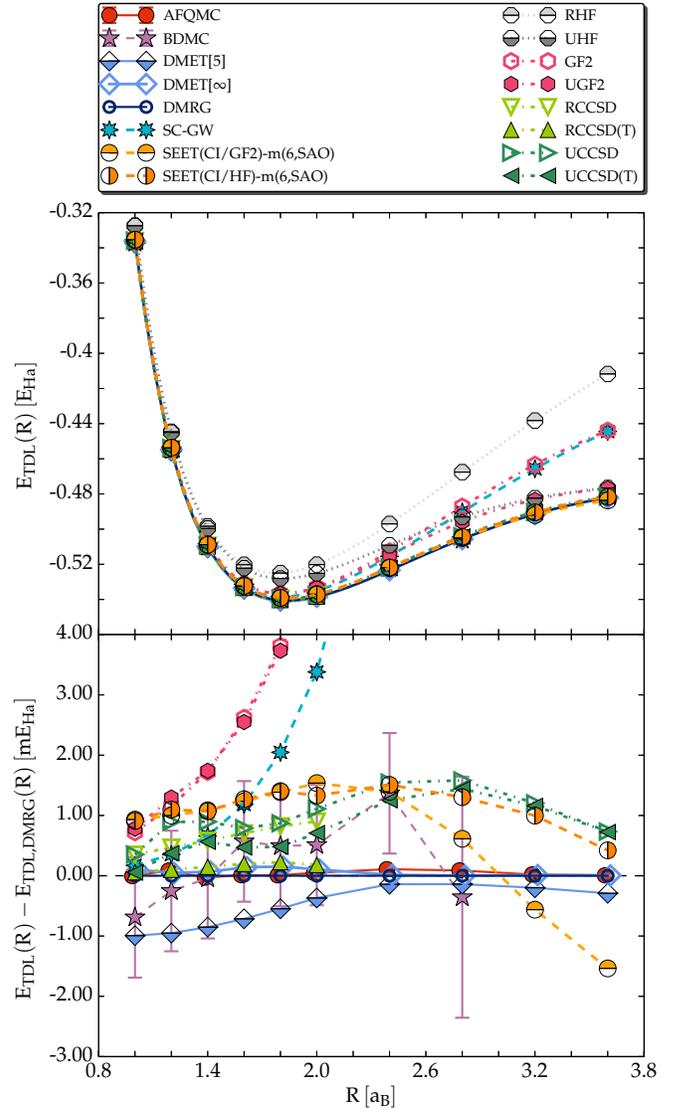}
\caption{
Top: Computed equation of state in the minimal basis at the thermodynamic limit.
Bottom: Detailed comparison using DMRG results as reference. 
}
\label{fig:tdl_sto}
\end{figure}

The minimal basis set hydrogen chain is similar to an extended Hubbard model. As such,
the results in this basis provide a quantitative connection to model studies.
The computed 
EOS is shown in Fig.~\ref{fig:tdl_sto} for the STO-6G basis set.
DMRG calculations can be carried out for large system sizes in this basis, and serve as 
a near-FCI quality benchmark. 
DMRG results for finite chains, after extrapolation to 
$N\to \infty$, yield an 
equilibrium geometry $R_e = 1.831(3)\,\bohr$ and ground-state energy per atom of $E_0 = -0.5407(2)\,\Ha$ at $R_e$. 

In the lower panel of Fig.~\ref{fig:tdl_sto}, a detailed comparison is shown using DMRG as a reference.
Most methods show similar behaviors as in finite chains. Coupled-cluster methods display the same general trends, 
 with RCCSD(T) in particular giving extremely accurate results before the breakdown 
 at larger bondlengths ($R>2\, \bohr$). 
 AFQMC yields energies accurate to within $0.15 \mHa$ per particle across the bondlengths. 

 SEET is extrapolated to the TDL with respect to the chain length, with the number of orbitals treated by an accurate method
fixed to 6. With this constraint, SEET(CI/HF)-m(6,SAO)  shows  accuracy at the thermodynamic limit comparable to the
10-atom chain when FCI is used to treat the impurity and HF to treat environment, with a maximum error of $\sim 1 \mHa$.
For stretched distances, SEET-m results improve if HF is used instead of GF2 since the latter (SEET(CI/GF2)-m(6,SAO)) results in overcorrelation.

 As discussed in Sec.~\ref{sec:dmet}, two types of DMET calculations were performed for the minimal basis.
 DMET[5] was from the first type, treating finite chains with fragment size $N_f=5$, followed by extrapolation of the chain
 size $N$ similar to the procedure used by most other methods 
  whose results are shown here; this gives a maximum error of $\sim 1 \mHa$.
 DMET[$\infty$] shows results from the second type, which worked directly in the large $N$ limit, 
 and extrapolated the fragment cluster size $N_f$. Details of the extrapolation procedure are given in Sec.~\ref{sec:dmet}. 
 DMET[$\infty$] results should approach the exact limit similar to 
 DMRG; the DMET[$\infty$] and DMRG energies agree to better than $0.15 \mHa$ per particle across all bondlengths.

BDMC$_3$ yields converged results up to $R = 2.4$. For $R=2.8$ convergence is reached
only at the level of BDMC$_5$; reaching convergence for larger values of $R$ requires even higher
orders. The calculated EOS is in good agreement with the exact results, and its final error bar 
of $1 \mHa$ for $R\leq 2.4$ is dominated by the resolution of the grid of 512 Matsubara frequencies used.
For $R=2.8$ the error bar of $2 \mHa$ is dominated by statistical noise in high diagrammatic orders.
(The performance of lower-order BDMC$_n$ calculations is discussed in Appendix \ref{sec:diagrammatic}, as well as their relations 
with SC-GW and other diagrammatic techniques.)

\subsection{Equation of state in the complete basis set and thermodynamic limits}
\label{sec:cbstdl}

Our final results for the equation of state of the hydrogen chain 
are presented in Fig.~\ref{tdl_cbs}. Detailed numerical data are tabulated and included in the Appendix. 
For these results, VMC and LR-DMC 
are extrapolated to the TDL, while 
basis-set methods are extrapolated to the joint TDL+CBS limit.
We carry out extensive self-consistency and cross-checks in order to validate the extrapolations, as 
discussed in Secs.~\ref{subsec:h10-extrap-x} and ~\ref{sec:tdl-extrp-x-N}.

\begin{figure}[h!]
\centering
\includegraphics[width=0.99\columnwidth]{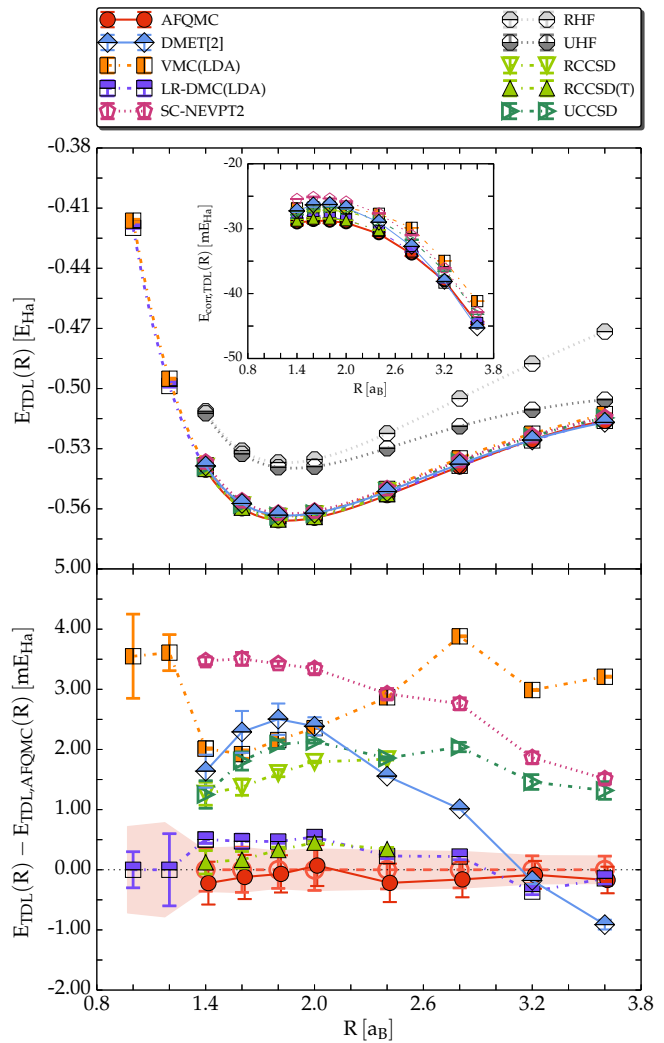}
\caption{
Top: Computed equation of state of the hydrogen chain in the thermodynamic limit.
The inset shows the corresponding correlation energy per particle. 
Bottom: Detailed comparison using AFQMC results as reference \REV{(LR-DMC for the shortest two bondlengths)}.
The empty circles indicate AFQMC results after a correction is applied from the difference with DMRG at the cc-pVDZ level. 
The pink error bands indicate all statistical uncertainties.
} \label{tdl_cbs}
\end{figure}

In the bottom panel of Fig.~\ref{tdl_cbs}, AFQMC results are used as a reference, based on its accuracy from 
the systems which have been benchmarked. Large system sizes and basis sets can be reached to minimize 
the uncertainty in the extrapolation to the TDL and CBS limits. We can further quantify 
the residual systematic errors of the constraint in AFQMC from cross-checks with DMRG, 
by estimating their difference, 
$E_{\mathrm{TDL,DMRG}}(R)-E_{\mathrm{TDL,AFQMC}}(R)$,
at the cc-pVDZ basis level. This ``correction'' can be applied to the AFQMC equation of state at the
CBS limit, $E_{\mathrm{TDL,AFQMC}}(R)$.
The result is shown by the empty circles and dashed lines in Fig.~\ref{tdl_cbs}, while original AFQMC data are 
shown with solid circles and lines. 

Agreement is seen between these results and that from RCCSD(T), which provides another consistency check.
We find an equilibrium bondlength of $R_{eq} = 1.859(3) \bohr$ with an energy of $E_0 = -0.5659(3) \Ha$
at $R_{eq}$ in the thermodynamic limit.
The computed correlation energy, defined with respect to the RHF energy, is shown in the inset of Fig.~\ref{tdl_cbs}. 

\REV
{
At the smallest bondlengths ($R \le 1.2$), there is significant linear dependence in the basis sets.
This causes an effective reduction in the size of the basis, which can render the usual
ansatz 
for basis set extrapolations unreliable. We thus avoid performing CBS extrapolations. 
(All finite-basis data are listed in the appendix and repository.
It will be valuable to develop specifically designed basis set sequences or correction methods in
this regime.)
LR-DMC results are shown, which provide an upper bound for the energy. 
Based on the results in Sec.~\ref{subsec:h10-CBS}, the fixed-node error is estimated to be
 $< 1\,\mHa$ per atom, which is indicated by the pink error bands on these two points.
}

LR-DMC results are obtained directly in real space, and provide an independent validation.
At large bondlengths, the fixed-node error in LR-DMC is minimal,  as we have seen in the finite-chain benchmarks.
Furthermore, we have performed PBC calculations using LR-DMC to provide a separate 
extrapolation to the TDL.
The excellent agreement between LR-DMC  and AFQMC at large $R$ is thus a further indication of the 
robustness of our procedures for reaching the infinite basis set and thermodynamic limits.
Note that the VMC results exhibit a different trend from the  corresponding LR-DMC, 
suggesting that the variational many-body wave function is best at intermediate bondlengths.
This is likely a reflection of the balance between the two parts that form the VMC ansatz, namely the LDA Slater determinant and the optimized  
Jastrow factor. The former becomes more accurate in describing the nodal surface as $R$ increases,
where the latter is evidently more effective at weaker correlation.
Only the  determinant part, via the nodes that it defines,  impacts the DMC results.

The DMET[2] results provide an example of an embedding calculation at the thermodynamic and complete basis set limits,
with a modest impurity size. The limitation on the impurity size is from the use of a DMRG impurity solver,
which becomes expensive in the large basis set limit.

We comment that various correction schemes can be applied to our finite-basis and/or finite-size 
data to provide additional estimates from methods not included in Fig.~\ref{tdl_cbs}. For example, a 
residual basis set correction could be obtained either from a different method or using a lower order
theory (if available) of the same method, and applied to DZ or TZ basis results to estimate the CBS limit. 
These can be  
readily retrieved for assessment from the detailed data provided in the appendices and supplementary materials.

\subsection{Reaching the complete basis set and thermodynamic limits}
\label{sec:tdl-extrp-x-N}

\begin{figure*}[t!]
\centering
\includegraphics[width=0.9\textwidth]{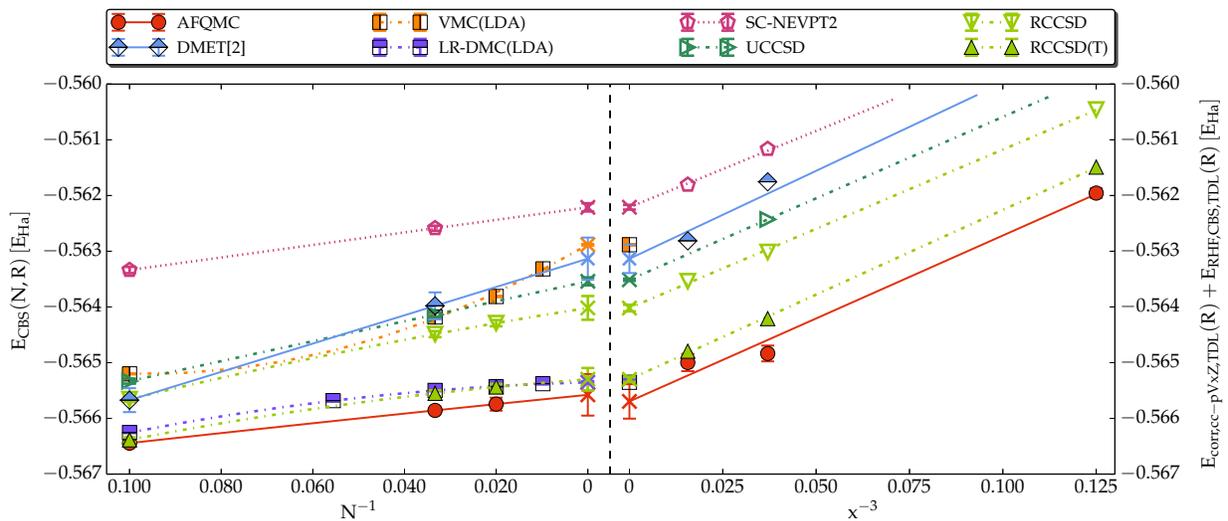}
\caption{
Illustration of the extrapolations to the CBS and TDL limits. 
Results are shown for $R=1.8 \bohr$. The left panel shows extrapolation of $E_{\rm CBS}(N)$ vs.~$1/N$,
while the right panel shows extrapolation of $E_{{\rm cc-pV}x{\rm Z}}(N\to \infty)$ vs.~$1/x^3$. (The correlation 
energy is shown on the right, shifted by the CBS RHF energy.) Final results are consistent within statistical errors 
and independent of the order with which the limits are taken.} \label{fig_correlated_tdlcbs} 
\end{figure*}

A key challenge in the \emph{ab initio\/} computation of bulk materials is to remove finite-size and 
finite-basis effects so as to obtain results for the continuous and thermodynamic limits. 
This is important in order to make reliable predictions about materials properties and allow direct 
comparisons with experiments.
Various choices exist in the calculation.
These can be at the level of the type of many-body methods, for example the use of particular
embedding approaches (versus those that treat a cluster only, whether finite or periodic), 
or the use of coordinate space methods like DMC versus basis set methods. 
They can also be common to classes of methods and decoupled from and independent of  the details of the underlying many-body methods, 
for example the use of periodic supercells versus finite 
clusters, or 
the choice of basis sets, etc. 
By employing many state-of-the-art methods, we are able to 
investigate these factors extensively and with great care in the present work.

Many of our calculations are performed using OBC,
i.e., treating a finite chain. 
We find that, somewhat surprisingly,  OBC calculations show faster 
convergence to the TDL than PBC \REVISED{in the hydrogen chain} for all but the smallest 
few bondlengths  \REVISED{(see Appendix~\ref{sec:extrapolation_technique}).} 
To extrapolate the finite-$N$ results to the TDL, 
we assume that the PEC has the following size dependence: 
\begin{equation}
\label{eq:tdl_xtrap}
E(N,R) = \sum_{i=0}^k \frac{ A_i(R) }{ N^{i} }\,, 
\end{equation}
where $k$ is a small integer.
For $k=1$, this gives the  subtraction trick based on a division of surface and bulk terms, namely  
$A_0 = \frac{ N_1 E(N_1) - N_2 E(N_2) }{ N_1 - N_2 }$ (omitting $R$), 
which has been 
used, for example, in DMRG calculations before \cite{Stoudenmire2012}.
In this work we typically used $k=2$, employing $N=10,30,50$ 
and, when necessary, $N=18$, $22$, $70$ and $102$.

Under this choice, there are still multiple strategies 
for finite basis set methods to approach the combined limits. One could extrapolate to the CBS limit
for each finite chain of fixed $N$ following the procedure 
described in Sec.~\ref{subsec:h10-extrap-x}, and then extrapolate the results in $N$ to 
the TDL.  Alternatively, one could extrapolate each basis set to the TDL, and then extrapolate to the CBS 
limit, or use a joint ansatz and extrapolate both simultaneously.
As illustrated in Fig.~\ref{fig_correlated_tdlcbs}, exchanging the order of the extrapolation leads to 
consistent and robust results.

With the exception of the minimal basis, the TDL extrapolation for the DMET data is performed using the same
OBC size-dependence described above. In the minimal basis, we also carried out DMET calculations
directly in the TDL, as mentioned earlier. Additional details on this extrapolation scheme and a comparison of the two is discussed 
briefly in the appendix.
Calculations in PBC (including those with BDMC, which used a ring geometry, and LR-DMC) 
are extrapolated to the TDL using the form $E(N,R) = A_0(R) + A_2(R) 
\, N^{-2}$  
\cite{Chiesa2006}, and statistical error bars are propagated following standard 
procedures in the extrapolations.

\section{Conclusions}
\label{sec:conclusion}

We have presented a comprehensive investigation of the hydrogen chain, deploying a vast 
suite of cutting-edge many-body numerical methodologies and obtaining a detailed and quantitative 
understanding of current computational capabilities for treating correlated quantum materials.
We have shown how finite-size effects and finite spatial or other basis set resolutions can be 
systematically removed to reach the physically relevant infinite system size and complete basis set limits.

Through the synergistic use of complementary methods, we have accurately determined the 
ground-state energy as a function of interatomic distance. 
This serves as a proof of concept for a new mode of attack on correlated materials by  \emph{ab initio} calculations.
The benchmark results will provide a reference on the state of the art in many-body computation of real materials. 

Our study captures many of the salient features of predictive computations in real materials. 
The ability of each many-body method to correctly capture important physical properties will depend on 
the material system under study. 
For example perturbative or diagrammatic methods can have better or worse accuracy in systems with 
different amount of electron correlation,
the qualities of the constraints on the sign problem in QMC methods can vary with the physical nature of 
the problem, the rate of convergence with the fragments or the requirement on the impurity 
solver  in embedding methods can differ from material to material, the scaling and computational 
feasibility of DMRG can change, etc.
 More benchmark studies of this kind will be highly desirable to broaden the understanding 
and identify further limitations as well as opportunities of development.

The computational cost of each method depends on various factors, including the degree to which the algorithm and 
codes have been optimized, the level to which one wishes to take the calculation
(the order in perturbative or diagrammatic methods, or the statistical accuracy in 
Monte Carlo methods), etc. 
The results  in this benchmark were obtained with moderate computing (order of days on
platforms ranging from local clusters to medium-sized supercomputers).
The computational 
scaling, which is summarized in Sec.~\ref{sec:methods}, together with the corresponding 
accuracy achieved by each method in the benchmark, will provide a rule of thumb on their computational 
cost.

The benchmark results indicate that many of the methods tested here are capable of reaching
an accuracy of five percent of the cohesive energy or better, across  wide parameter  regimes of strong electron correlation. 
A subset of these methods predict the equation of state systematically to sub-milli-Hartree accuracy.
Further development may turn these into post-DFT methods of choice for ground-state studies,
\REVISED{and a concerted effort to build open-source codes will be invaluable.}
Other techniques can more naturally address dynamical and thermodynamical properties, many of which 
are the outcome of recent research.
Continued development along these lines will further improve their accuracy and time to solution.
Further benchmark studies of dynamical and thermal effects, building on the work done here on the equation
of state, would also be very desirable.

\REVISED{It is important to continue to expand the benchmark studies to more complex materials.}
Even in this relatively simple system of the hydrogen chain, important questions remain on the physics which are of strong interest 
and relevance to some of the key issues in strongly correlated systems in general. 
For example, how does the nature of the charge and magnetic orders vary with the bondlength? 
We are presently investigating these and related questions.

\begin{acknowledgments}

We gratefully acknowledge the Simons Foundation for funding.
We thank E.~Kozik, H.~Krakauer, M.~van Schilfgaarde, H.~Shi, B.~Svistunov and N.~Tubman for valuable interactions.
Support from NSF (Grant no.~DMR-1409510) is acknowledged for method development work at 
William \& Mary.
F.~M.~was also supported by DOE (Grant no.~DE-SC0001303).
The work at the California Institute of Technology was 
supported by the Department of Energy, through DOE-SC0008624. G.~K.-L.~C. is a Simons Investigator.
The work at Rice University was supported by Grant
No. NSF-CHE-1462434. J.~A.~G. acknowledges support from the National Science Foundation Graduate 
Research Fellowship Program (DGE-1450681). G.~E.~S. is a 
Welch Foundation Chair (C-0036).
I.~S.~T. and N.~V.~P. acknowledge NFS under the grant PHY-1314735.
S.~S. acknowledges computational resources provided through the HPCI System Research Project (Nos. hp160126) 
on the HOKUSAI GreatWave computer under project G16026.
S.~R.~W. and E.~M.~S. acknowledge support from the U. S. Department of Energy, Office of Science, Basic Energy Sciences under award \#DE-SC008696.
E.~G. was also supported by DOE Grant No. ER 46932, J.~L. by NSF DMR 1606348, and computer resources were provided by TG-DMR130036.
D.~Z. and T.~N.~L. were also supported from DOE Grant no. ER16391.

\end{acknowledgments}

\pagebreak

\appendix

\section{Description of computational methods}
\label{sec:methods-description}

In this Appendix, we provide further descriptions of the  individual methods used.
Following the main text, we group methods into the following categories: deterministic wavefunction (CC, 
DMRG, MRCI, NEVPT2), stochastic wavefunction (AFQMC, LRDMC, VMC), embedding (DMET, SEET) and
diagrammatic (SC-GW, GF2, BDMC).
The categories are by no means rigid, as a method can fit into multiple groups; 
they are meant to provide a general guide and help with organization of the discussions.

\subsection{Deterministic wave function methods}

Deterministic wavefunction methods (HF, CC, DMRG, FCI, MRCI, NEVPT2) range in quality between the mean-field HF and the exact FCI. 
Correspondingly, their computational costs vary a great deal. 
These methods rely on different types of ansatz, the nature of which is ultimately responsible
for their accuracy and computational cost.

\subsubsection{Coupled Cluster (CC)}

Coupled cluster (CC) theory \cite{Paldus1999,Bartlett2007,Shavitt2009} is 
widely applied, 
often providing accurate and systematically-improvable ground
state energies when systems are neither too large nor too strongly correlated. The CC wavefunction
is written as
\begin{equation}\label{Eq:exp}
\ket{\mathrm{CC}} = e^{\hat{T}}\ket{0},
\end{equation}
where $\ket{0}$ is a single-determinant reference state, and
the cluster operator is given by
\begin{equation}\label{Eq:T}
\hat{T} = \sum\limits_{\mu} t_{\mu}\hat{t}_{\mu},
\end{equation}
where $\hat{t}_{\mu}$ creates an excited determinant $\ket{\mu}$ containing
$\mu$ particle-hole pairs relative to
$\ket{0}$, with amplitude $t_{\mu}$.
Standard CC theory constructs a similarity-transformed Hamiltonian
\begin{equation}\label{eq:hbarcc}
\hat{H}^\prime = e^{-\hat{T}} \hat{H} e^{\hat{T}},
\end{equation}
and the energy and amplitudes $t_{\mu}$ are obtained by solving the Schr\"odinger equation
projectively
\begin{subequations}\label{eq:CC}
\begin{align}
E &= \braket{0| \hat{H}^\prime |0} ,\\
0 &= \braket{\mu| \hat{H}^\prime |0} & \forall \mu.
\end{align}
\end{subequations}
In other words, CC theory diagonalizes the similarity transformed Hamiltonian in the space spanned
by the mean field reference and excitation manifold. Because $\hat{T}$ is an excitation operator,
the commutator expansion used to evaluate the amplitudes in \ref{eq:CC} terminates after four
nested commutators for all values of $\mu$, because the Hamiltonian contains only one-
and two-particle terms.
In this work, we limit $\hat{T}$ to $\mu \leq 2$, i.e. CC with single and double excitations (CCSD) on
restricted
Hartree-Fock (RCCSD) and unrestricted Hartree-Fock (UCCSD) references.\cite{Purvis1982,Scuseria1987}
Since the quality of the reference
determines the quality of the CC wavefunction, a reference obtained from a symmetry-broken mean field can be critical to getting
good CC energies when systems become strongly correlated. When multiple Hartree-Fock solutions exist,
the question of which one to use as a reference for the CC calculations can be subtle. In general,
our CC calculations are performed
on the lowest-energy Hartree-Fock determinants we could find. In some calculations, we have also
perturbatively included triple-excitation
effects, denoted CCSD(T) \cite{Pople1987}. 

All CC calculations shown here
are widely available in standard
quantum chemistry packages~\cite{g09d1}.
To carry out the larger calculations, we used the high-performance implementation in the \textsc{PySCF} package.
This uses an AO-driven implementation to reduce the IO costs associated with accessing integrals on disk~\cite{koch1994direct}.
We carried out RCCSD and RCCSD(T) calculations for H$_{30}$ with cc-pVQZ and cc-pV5Z basis and H$_{50}$ 
with cc-pVTZ and cc-pVQZ basis with this implementation.
Even with the AO-driven technique, the RCCSD calculation for
H$_{50}$ in the cc-pV5Z basis would require at least 4 TB of disk space. Although technically feasible,
we did not perform it here.
For the largest basis sets cc-pVTZ, cc-pVQZ and cc-pV5Z at geometries $R <1.6 \bohr$ and cc-pVDZ at $R < 1.4 \bohr$, 
the Gaussian basis was nearly linearly dependent.
We removed  linearly dependent vectors with an overlap
threshold of $10^{-7}$.  Although a smaller threshold could be used in the Hartree-Fock
calculation, changing the energy by $\sim 10^{-6}$ $\Ha$ per atom, we found the resulting CCSD calculations with
smaller thresholds to be numerically unstable.

\subsubsection{Density Matrix Renormalization Group (DMRG, SBDMRG)}
\label{sec:dmrg}

DMRG (density matrix renormalization group) is a low-entanglement wavefunction approximation~\cite{white1992density}. The wavefunction can
be written as 
\begin{align}
  |\Psi\rangle = \sum_{ \{ n\}} \mathbf{A}^{n_1} \mathbf{A}^{n_2} \ldots \mathbf{A}^{n_M} |n_1 n_2 \ldots n_M \rangle
  \end{align}
where $M$ denotes the number of orbitals in the system.
Each $\mathbf{A}^{n}$ is a $D \times D$ matrix
of real numbers associated with a single-particle basis function,
except for the boundary terms $\mathbf{A}^{n_1}$ and $\mathbf{A}^{n_M}$, which are length-$D$ vectors. $D$ denotes the bond dimension and
controls the accuracy of the simulation; as $D$ increases, the
wavefunction converges to the exact correlated state.
In linear systems such as the hydrogen chains considered here, provided that the gap of the system does not close,
the bond dimension required for a given accuracy per atom stays close to a constant, independent of the number of atoms.
The energy of the wavefunction may be stably computed and variationally optimized through the DMRG sweep algorithm.

In this work we considered two different kinds of single-particle basis functions
in the DMRG calculations. In the standard quantum chemistry formulation of DMRG~\cite{white1999ab,chan2002highly,chan2016matrix}, the 
single-particle basis is simply an orthogonal basis in the space of Gaussian orbitals of the system. This is what we will refer to as 
DMRG in the calculations in this work, and details can be found in standard references to
DMRG in the quantum chemistry literature~\cite{dmrgqc}. 
We carried out Gaussian based DMRG calculations using the implementation in the \textsc{Block} code,
with the standard settings described in Ref.~\cite{olivares2015ab}. 
DMRG energies were computed for H$_{10}$ (STO-6G, cc-pVDZ, cc-pVTZ bases), H$_{30}$ (STO-6G, cc-pVDZ),
and H$_{50}$ (STO-6G),  up to a maximum bond dimension of $D=2000$. For the STO-6G basis, the DMRG single-particle basis
was the set of symmetrically orthogonalized AO orbitals. For all the other bases, we used split-localized molecular orbitals.
The split-localized orbitals were ordered by the exchange Fiedler vector~\cite{barcza2011quantum,olivares2015ab}.
The estimated maximum uncertainty in the Gaussian based DMRG energies is less than $0.05$ $\mHa$ per atom.

In addition, we have also \REVISED{introduced} the sliced basis DMRG method (SBDMRG).
Here, instead of a 3D Gaussian basis, one uses a grid in one direction~($z$), while using a
Gaussian-derived basis in the two transverse directions~($x$,$y$). Formally,
one can write the basis as 
\begin{align}
\phi_{j n}(x,y,z) = \varphi_{j n}(x,y) \, \delta^{\frac{1}{2}}(z-z_n) \ , \label{eqn:sbdmrg_basis}
\end{align}
where $n$ labels grid points along the $z$ direction (with grid spacing $a$) and $j$ labels the
transverse basis function at that grid point (or ``slice'').
The $\frac{1}{2}$ power on the Dirac delta function indicates that
the basis functions are square-normalized. 
The kinetic energy terms in the Hamiltonian are approximated with a fourth order finite-difference second 
derivative approximation. For the data presented in Section~\ref{sec:cbstdl}, 
we used a  grid spacing of \mbox{$a=0.1$}, for which we estimate an error of about $0.1 \, \mHa$ per atom.
The transverse basis functions $\varphi_{jn}(x,y)$ are derived from a standard Gaussian, atom-centered basis
set. Functions from the standard basis are projected onto each slice, then these functions are orthogonalized,
keeping the most significant ones up to as many as the number of contracted orbitals on each atom.
Compared to the original basis, a sliced basis has approximately the same transverse resolution, 
but its resolution in the $z$ direction is essentially perfect. Thus energies in the sliced basis are generally
lower than in the original basis, due to the improved $z$ correlation.
The key advantage of SBDMRG over the standard quantum chemistry formulation of DMRG is that 
the local support of the basis functions along the grid direction makes the number of Hamiltonian terms 
proportional to $M^2$.
Using matrix product operator compression techniques, which are quite simple to apply
in the sliced basis formulation, the cost of SBDMRG is further reduced to $M$, 
which is the same scaling as applying DMRG to the 1D Hubbard model. 
For more details see the recent paper Ref.~\onlinecite{Stoudenmire:2017}.
 
\subsubsection{Multireference configuration interaction (MRCI)}

MRCI (multireference configuration interaction) is a method that incorporates 1- and 2-external
excitations on top of an active space wavefunction. It is a commonly used method for
high accuracy simulations of multireference electronic structure in small molecules.
Here we use a variant of internally contracted
MRCI described by Werner and Knowles~\cite{werner1988efficient,knowles1988efficient}, implemented in the \textsc{Molpro} package~\cite{werner2012molpro}. We
start with a CASSCF (complete active space self-consistent-field) wavefunction $|\Psi_0\rangle$. The variational ansatz is then
\begin{align}
|\Psi\rangle = c_0 |\Psi_0\rangle + \sum_I c_I | \Psi_I\rangle + \frac{1}{4} \sum_{ijab} c_{ij}^{ab} \hat{E}_i^a \hat{E}_j^b | \Psi_0 \rangle
\end{align}
where $|\Psi_0\rangle$ is the CASSCF wavefunction, $|\Psi_I\rangle$ is a configuration state function (CSF) with
a single external orbital, and $\hat{E}_i^a$ is the spin-summed excitation operator, $\sum_{\sigma} \hat{a}^\dag_{a\sigma} \hat{a}_{i\sigma}$.
The parameters $c_0$, $c_I$, and $c_{ij}^{ab}$ are determined variationally.

We also considered the explicitly
correlated (F12) MRCI approximation~\cite{shiozaki2011explicitly,shiozaki2013multireference,shiozaki2011explicitlyB}. Explicit correlation  accelerates convergence to 
the complete basis set limit by introducing 2-external amplitudes with explicit $r_{12}$ dependent functions. The associated integrals are computed through an auxiliary 
Gaussian basis. In this work, we used the default F12 settings and auxiliary bases in \textsc{Molpro}, including the singles corrections in the complementary auxiliary 
basis set space.

The MRCI wavefunction does not give an extensive energy. Defining the MRCI correlation energy as $\Delta E = \langle \Psi | \hat{H} | \Psi\rangle - \langle \Psi_0 | \hat{H} |\Psi_0\rangle$, 
we define the approximate size-extensive correlation energy (Q) through the scaling  $\Delta E \to \Delta E (1-c_0^2)/c_0^2$.

With the above techniques, we computed MRCI+Q and MRCI-F12+Q wavefunctions and energies for H$_{10}$ in the STO-6G and cc-pV$x$Z ($x$=2-5) bases, using a 
(10, 10) CASSCF initial state.

\subsubsection{$N$-electron valence state perturbation theory}

$N$-electron valence state perturbation theory (NEVPT) \cite{angeli2007new} is a multireference 2nd order
perturbation theory which is size-extensive and free of intruder state problems. It uses a zeroth order CASSCF wavefunction
and Dyall's Hamiltonian~\cite{dyall1995choice} as the zeroth order Hamiltonian. From this starting point, the 1st order wavefunction and
 2nd order energy are defined using the usual Rayleigh-Schr\"odinger perturbation theory. Typically
the 1st order equation is not solved exactly, but rather in a restricted variational space. In partially-contracted NEVPT2 (PC-NEVPT2), the
1st order wavefunction is expanded in the space of 1- and 2-external excitation operators acting on the ground-state wavefunction. 
In strongly-contracted NEVPT2 (SC-NEVPT2), the expansion space of the 1st order wavefunction is restricted further such that the amplitudes can be
determined without solving any linear equations, and simply from expectation values of the zeroth order wavefunction. 

For the   H$_{10}$ chain, we computed SC-NEVPT2 and PC-NEVPT2 using the \textsc{Molpro} package, starting from a (10, 10) CASSCF
zeroth order state. 
For H$_{30}$ we carried
out  DMRG-SC-NEVPT2 calculations~\cite{guo2016n} using the \textsc{PySCF} package, starting from a (30, 30) DMRG-CASSCF zeroth order state computed with split-localized orbitals, using the \textsc{Block} package. The basis linear dependency threshold was set to $10^{-8}$. The DMRG-CASSCF calculation was carried out with bond dimension 
$D=1000$, leading to an estimated energy error of less than $0.1 \mHa$. The zeroth order wavefunction was constructed by 
 compressing the DMRG-CASSCF wavefunction down to bond dimension $D=500$, with a compression error in the
total energy of less than $0.3 \mHa$
except at the shortest geometry, $R=1.0 \, \bohr$, where it was $10 \mHa$.
The semi-internal components of the DMRG-SC-NEVPT2 wavefunction and energy were approximated using the MPS compression scheme described in~\cite{sharma2014communication,sharma2017combining}, with a first order wavefunction bond dimension of $D=1500$; these
contributions were determined with an estimated accuracy of $0.1 \mHa$.

\subsection{Stochastic wave function methods}

Stochastic wavefunction methods (AFQMC, VMC, LRDMC) rely on Monte Carlo sampling
to construct an ansatz for
the ground state of the system and compute expectation values of observables.
AFQMC and LRDMC are both based on mapping the
imaginary-time evolution onto a random walk. AFQMC is formulated in non-orthogonal 
Slater determinant space. LRDMC (and VMC) conducts the random walk in coordinate space.
The fermion sign problem has different manifestations in the different manifolds, and the constraints
to control them lead to different approximations.

\subsubsection{Auxiliary-field quantum Monte Carlo (AFQMC)}

The auxiliary-field quantum Monte Carlo (AFQMC) is a wavefunction method, which estimates the ground-state properties of a many-fermion system
by statistically sampling the wavefunction 
$e^{-\beta \ham} \ket{\Psi_0} \propto \ket{\Psi_\beta} \rightarrow \ket{\Psi_G} $,
where $\Psi_0$ is an initial wavefunction, non-orthogonal
to the ground state $\Psi_G$
of $\ham$
\cite{Zhang1997,Zhang2003}.
For sufficiently large $\beta$, expectation values computed over
$\Psi_\beta$ gives ground-state averages.
AFQMC projects $\Psi_0$ towards $\Psi_G$
iteratively, writing $e^{-\beta \ham} = (e^{-\delta\beta \ham})^n$ where
$\delta\beta = \frac{\beta}{n}$ is a small imaginary-time step.
The propagator is represented as $e^{-\delta\beta \ham} = \int d{\bf{x}} \, p({\bf{x}}) \, \hat{B}({\bf{x}})$,
where $\hat{B}({\bf{x}})$ is a mean-field propagator of the form of an exponential of one-body operators that are dependent on 
the vector ${\bf{x}}$, and $p({\bf{x}})$ a probability distribution \cite{Stratonovich1957,Hubbard1959}.
This representation maps the original interacting system onto an ensemble of non-interacting systems subject to a fluctuating
potential. 
The imaginary-time projection can be realized as an open-ended random walk over the auxiliary-field (i.e., mean-field potential) configurations \cite{Zhang1997,Zhang2003}.
Sampling the trajectories of the random walk leads to a stochastic representation of ${\Psi_\beta}$ as an ensemble 
of Slater determinants.

For general two-body interactions AFQMC has a sign/phase problem, which is controlled by a phaseless gauge constraint (CP) on the Slater determinants using a trial wavefunction ${\Psi_T}$ \cite{Zhang2003,Zhang-lectureNotes}.
(For Hamiltonians that satisfy certain symmetry properties, e.g. the Hubbard model at half-filling, AFQMC is free of the sign problem).
The trial wavefunction is typically taken from Hartree-Fock or DFT. 
A self-consistent constraint is possible \cite{Qin-SC-CPMC-PhysRevB.94.235119} but is not 
used in this work. 
The accuracy of the CP AFQMC was extensively benchmarked in both real materials \cite{AlSaidi2007,
Purwanto2009,Virgus2014}
and lattice models \cite{LeBlanc2015,Qin-SC-CPMC-PhysRevB.94.235119}.
The CP AFQMC provides an alternative and complementary way to addressing the sign problem with respect to fixed-node DMC.
The random walks take place in the overcomplete manifold of Slater determinants, in which fermion antisymmetry is by construction maintained in
each walker. Applications have indicated that often this reduces the severity of the sign problem and, as a result, the phaseless
approximation has weaker reliance on the trial wave function \cite{Purwanto2008}.

For \emph{ab initio\/} materials computations, AFQMC can be carried out using either a plane-wave
basis and pseudopotentials \cite{Zhang2003,Ma2016}, localized basis sets such as 
standard Gaussian type orbitals  
 \cite{Al-Saidi-Gaussian-AFQMC-doi:10.1063/1.2200885}, or general basis sets using DFT orbitals
 \cite{Ma2015}. 
In this work, we apply AFQMC implemented for Gaussian basis sets to finite chains.
In all our calculations we use a linear dependence threshold of $10^{-7}$ in the one-electron basis.
The two-body matrix elements $v_{pqrs}$ are decoupled into bilinear form with the modified Cholesky approach 
using a tolerance of $\delta \leq 10^{-5}$ \cite{Purwanto-JCP-2011}. 
Results are extrapolated to the TDL and CBS limits.
The total projection time is typically $\beta = 80 \, \Ha$, although calculations
with $\beta = 220 \, \Ha$ were performed in some cases.
The convergence error from the use of a finite $\beta$ is negligible.
Most calculations
used $\delta \beta = 0.005 \, \Ha^{-1}$.
Extrapolations were performed where the associated Trotter error was greater than the other uncertainties.
The reported QMC error bars are estimated as one standard deviation statistical errors.
The CP bias leads to a non-variational estimator of the ground-state energy.
In our calculations, the UHF ground state was taken to be $\Psi_T$.
For the two largest bondlengths ($R=3.2$ and $3.6$), motivated by the analogy between the \chem{H} chain and 
the Heisenberg model, we employed \REVISED{in this work a new form of trial wavefunction}, a linear combination of spinon excitations on top of
the UHF state, $|\Psi_T \rangle = \sum_{k=0}^s \sum_{i_1<\dots<i_k} C_{i_1 \dots i_k} 
|\Psi_{i_1 \dots i_k} \rangle$.
Spinon excitations $\Psi_{i_1 \dots i_k}$ are constructed using atomic positions with antiferromagnetic
spin ordering, and $k$ pairs $(i_1, i_{1}+1) \dots (i_k,i_k+1)$ of adjacent spins flipped, as initial conditions 
for the UHF self-consistence procedure, and coefficients $C_{i_1 \dots i_k}$ are finally optimized variationally.
We used $k=1,2$ for $R=3.2$, $3.6$ respectively.

\subsubsection{Variational and Lattice-Regularized Diffusion Monte Carlo (LR-DMC, VMC)}

The LR-DMC is a projection method \cite{lrdmc} that uses the lattice regularization 
for applying the imaginary time propagator $\exp( -\beta \hat{H})$ to a trial function $\Psi_T$ defined in the continuous space.
\REVISED{In this work, improvements including new use of localized basis sets were introduced,
which drastically improved accuracy over previous results \cite{Sorella2011}.}
The main approximation is to write the Laplacian by means of its dicretized expression on a lattice with  a grid with lattice space $a$, e.g. for a single electron 
wavefunction depending only on one variable $x$:
$\nabla \Psi = \partial_x^2 \Psi \to \nabla_a \Psi = {1\over a^2} ( \Psi( x+a)+ \Psi(x-a)-2 \Psi(x))$. 
The extension of $\nabla_a$ to higher dimensions is 
 straightforward and the continuous 
space can be sampled ergodically even for finite $a>0$ with a much simpler 
algorithm than the original proposal \cite{lrdmc}, namely 
by randomizing the directions along which 
the Laplacian is discretized \cite{lrdmcsize}.
For $a\to 0$ and $\beta \to \infty$ the exact ground state wavefunction 
$\Psi_G$ can be obtained if $\langle \Psi_T | \Psi_G \rangle \ne 0 $.
 However, due to the fermion ''sign problem'', an  approximation is 
employed to achieve small and controlled statistical errors: it is required 
that, during the projection,  the ''sign'' of the propagated wavefunction 
is constrained to the one of the chosen trial function:
\begin{equation} \label{fixednode}
\Psi_T(x) \times \langle x | \exp( - \beta \hat{H} ) |\Psi_T\rangle \ge 0
\end{equation} 
for any electronic configuration $x$ where the spins and the electron positions are defined.
For $a\to0$ the results coincide with the standard Fixed-Node approximation introduced long time ago \cite{reynolds1982}.  This scheme  is usually employed 
 within the diffusion short time ($\Delta$) approximation of the  
propagator, in a way 
 that  in the   $\Delta \to 0$ limit, by applying it $\beta\over \Delta$ times,  the exact Fixed-Node projection scheme  (\ref{fixednode}) is recovered, with the well 
 established variational property on the estimated energy. 
In this work we have used the lattice regularization, just because it is more conveniently implemented in the TurboRVB package \cite{TurboRVB}, and also 
because  the extrapolations for $a\to 0$ are very well behaved and easily controlled in an automatic way.

This method is very weakly dependent on the dimension of the basis set 
chosen to represent the nodes of the Slater determinant, and we have 
verified that a negligible error in the DMC  energy is 
obtained by using  the standard cc-pVTZ  basis where the largest $Z_{1s}$ ($Z_{1s}=33.87$) is removed. Indeed too large exponents are also not necessary in this approach  
because the cusp conditions are fulfilled by a one-body Jastrow factor of the type:
\begin{equation}
 u_{1-body}(r) =- { 1 -\exp( -\sqrt{2} r) \over  \sqrt{2}}.  
\end{equation}
 This one body Jastrow is included also in the GTO basis set for the DFT (LDA) 
calculation. The DFT is  also defined (within the TurboRVB package \cite{TurboRVB}) on a mesh of lattice space $a=0.1$ or smaller, until convergence is reached 
within $0.001 \Ha$ in the total energy. The use of the one body Jastrow factor drastically improves the convergence for $a\to 0$ and the quality of the basis set as 
the DFT energy is much lower than the standard one in the original 
cc-pVTZ basis. 
\REVISED{For $R<1.4$ we found that the cc-pVDZ basis can be significantly
improved by adding $p$ diffusive Gaussian orbitals with small exponents
($Z_{1p}=0.2$, $Z_{2p}=0.05$), allowing us  to obtain the best
variational LR-DMC estimates for $R=1.0$ and $R=1.2$, even
better than with the  larger cc-pVTZ basis.}
Within periodic boundary conditions in the direction of the chain, assumed to be along the $z$ direction, we use a supercell of dimension 
$ L_x \times L_x \times L_z$ with $L_x=L_y= 40 \bohr$, that is large enough 
for safely neglecting  the interaction between the periodic images in the $x,y$ directions  (error less than  $0.0001 \Ha$ per atom).
Moreover the basis set (standard for open systems) has been periodized 
according to 
the standard procedure described in [\onlinecite{crystal}], when PBC are applied.

Before the application of the LR-DMC algorithm, a more accurate Jastrow 
factor is used to define the trial function $\Psi_T$. This contains the so-called two-, three- and four-body contributions that are expanded in a localized basis different 
from the determinant one. All these terms are efficiently optimized, using the scheme described in \cite{zenwater}.
Since the Jastrow is not affecting the results for $a\to 0$ presented here, 
we do not describe in details its form and the standard 
optimization methods used \cite{marchimol}.

\subsection{Embedding theories}

Quantum embedding theories (DMET, SEET) are based on the idea of combining two different types of quantum calculations: 
high-level calculations on one or more active regions of interest, called fragments, and low-level calculations on the 
environment surrounding fragments. 
In various methods, these fragments can be chosen either in the energy or in the local basis.

A quantum embedding theory determines the coupling between fragments and environment self-consistently,
using a variable of interest to provide for feedback.
DMET and SEET respectively use the one-particle density matrix and the self-energy as variables of interest.

\subsubsection{Density Matrix Embedding Theory (DMET)}
\label{sec:dmet}

DMET \cite{Knizia-2012,Knizia-2013} provides a framework to approximate
expectation values of a large system from embedded calculations. A
mean-field wavefunction $\Phi $ over the full system is used
to define the embedding of a fragment defined in terms of a set of $L$
local orbitals. The embedding of $\Phi $ splits the occupied
and virtual orbitals into ones with and without weight in the local
fragment. The occupied and virtual set {\em with} weight in the
fragment fully span the $L$ local orbitals used to define the
embedding, as well as an additional set of (at most $L$) {\em bath}
orbitals. In DMET, a high level calculation is carried out in the
fragment + bath orbital space; as the size of the embedded fragment
$L$ approaches that of the full system, the resulting DMET energy
converges to that of a high level calculation on the full
system. While DMET can be used to study finite systems, it provides a
natural framework to study systems directly in the TDL.  In this work,
we have used different DMET strategies for calculations in finite
chains and in the TDL; we refer the reader to Section B3 for further
details about the latter.

As outlined above, the splitting of the full system into fragments
requires the introduction of a set of local orbitals.  In this work,
we use intrinsic atomic orbitals (IAOs) \cite{Knizia-2013b} to define
the local basis in the valence space. In our finite chain
calculations, we split the system into fragments of $x$ atoms by
considering the corresponding local valence orbitals. These local
orbitals, through the embedding construction, generate a set of bath
orbitals. To this embedding (fragment + bath) space, we further add a
set of local virtual orbitals, built as projected atomic orbitals
(PAOs), on the constituent atoms. We use the acronym DMET[X] for
calculations using fragments of size X$=2,5$.

Expectation values in DMET (such as the energy and particle number)
are computed by partial traces \cite{Wouters-2016} (using the local
orbitals in a given fragment) of the contraction of integrals with
density matrices. A self-consistency loop can be used to uniquely
define $\Phi $ \cite{Knizia-2012,Wouters-2016}. In this work,
however, we use the RHF wavefunction without further optimization. A
global chemical potential is used to control the total number of
electrons in the system.

Our DMET results are reported using FCI as a solver for STO-6G basis
calculations and larger basis calculations with fragments of size
2. Other calculation use DMRG as a solver using a bond dimension of
$D=1000$. The error due to the DMRG solver is expected to be
significantly smaller than the error due to the fragment sizes
considered.

\REVISED{We have introduced new schemes for basis set and TDL
extrapolations in this work, further details for which} are provided in 
Section \ref{sec:extrapolation_technique}.

\subsubsection{Self-energy embedding theory (SEET)}

The self-energy embedding theory (SEET)~\cite{AlexeiSEET,LAN,nguyen2016rigorous,Zgid_Gull_seet_general,lan_mixing} relies on the assumption that orbitals in the 
system can be separated into $S$ different intersecting or non-intersecting subsets $A_i$, each containing $M^A_i$ orbitals, while  $M^R$ orbitals are contained in the the remainder $R$ 
such that $M^A_i \ll M$ for each $i$. 
In SEET, orbitals within subsets are strongly correlated and treated non-perturbatively; on the other hand, inter-subset correlations can be treated either perturbatively
or non-perturbatively.
In a case of a perturbative treatment, the inter-set correlations between  two different orbital sets $A_i$ and $A_j$, where $i\neq j$, or correlations in the remainder $R$
 are assumed to be weaker. 
Various ways of choosing the orbital subsets are possible.
\REVISED{New versions implemented in this work have lead to much improved accuracy.
We} employ a selection based on  the occupancies of natural orbitals (NOs) as well one based 
on the spatial locality of symmetrically  orthogonalized atomic orbitals (SAOs), and localized molecular orbitals (LMOs).
For details concerning each procedure see \cite{nguyen2016rigorous, lan_mixing}.

In SEET, the solution of the whole physical system is approximated by an affordable but frequently not so accurate $\Phi$-derivable method suitable for treating weakly 
correlated systems. Subsequently, this approximation is corrected within chosen strongly correlated orbital subspaces by a non-perturbative method. 
We have demonstrated that the general SEET functional can be written as
\begin{eqnarray}\label{eq:seet_mix_func} 
\Phi^\text{SEET}_\text{MIX}=&\Phi^{tot}_{weak}+\sum^{\binom{N}{K}}_{i}(\Phi^{A_i}_{strong}-\Phi^{A_i}_{weak})  \\ \nonumber
&\pm\sum_{k=K-1}^{k=1}\sum^{\binom{N}{k}}_{i}(\Phi^{B^k_i}_{strong}-\Phi^{B^k_i}_{weak}),
\end{eqnarray}
where the contributions with $\pm$ signs are used to account correctly for the possible double counting, for details see \cite{lan_mixing}. 
In this paper, $\Phi^\text{tot}_{weak}$ is determined from GF2 or HF.
In general, other choices such as  the GW method~\cite{Hedin1965} are also possible.
$\Phi^{A_i}_i$ stands for all those terms in $\Phi$ with all four indices $i,j,k,l$  of two-body interactions $v_{ijkl}$ contained inside orbital subspace $A_i$.
Here, $\Phi^{A_i}_{weak}$ is the solution for subset $A_i$ within the weak-coupling method, here GF2. $\Phi^{A_i}_{strong}$ is the solution in the $M_i^A$ subspace evaluated 
using a higher-order method suitable for treating ``strong correlation''.
We denote this way of performing SEET calculations as SEET(method strong/method weak)-m([$M^{A}$o]/basis) since here self-energies from intersecting orbital subspaces with $M^{A}$ strongly correlated orbitals are ``mixed'' between each other. While in a general case, the self-energy has to be evaluated for ${M}\choose{M^A}$ orbital subgroups, where ${M}\choose{M^A}$ can be a fairly large number, in practice one can quite trivially reduce it by identifying most important subgroups containing $M^A$ orbitals that lead to the significant lowering of the ground state energy, see \cite{lan_mixing}.

Since calculations can be performed either in the energy or spatial basis employing  NOs or SAOs and LMOs, respectively, as basis functions, we denote these choices using the ``basis'' keyword, where basis=NO, SAO, or LMO.

In SEET, the self-energy is constructed as a functional derivative of the $\Phi^\text{SEET}$-functional and the total SEET self-energy contains diagrams from both the
`strong' and `weak' coupling methods.

Consequently, each strongly correlated subspace self-energy is embedded into a  weakly correlated self-energy generated by all orbitals outside the strongly correlated subspace and accounting for all the non-local interactions on the strongly correlated orbital groups~\cite{Rusakov14}. For details explaining how to evaluate $\Sigma^{A_i}_{strong}$ and $\Sigma^{A_i}_{weak}$, we refer the reader to \cite{AlexeiSEET,Zgid_Gull_seet_general}.

We converged the electronic energy to $10^{-4} \Ha$.
The inverse temperature $\beta$ was set at 100 $\Ha^{-1}$ or 200 $\Ha^{-1}$ depending on the geometry.
The Matsubara freqency grid was generated using the splines interpolation \cite{Kananenka16} with the maximum number of points varying between 20,000 and 50,000.
In this paper, for the weakly correlated method in SEET, we used GF2 as well as HF. The strongly correlated part of the SEET self-energy was evaluated from the Anderson Impurity  model using FCI or versions of restricted active space CI (RASCI).
Note that in all the calculations presented here, SEET is based on the RHF reference.

\subsection{Diagrammatic methods}
\label{sec:diagrammatic}
 
The diagrammatic methods discussed in this work (SC-GW, GF2, BDMC)  evaluate, either deterministically (GF2, SC-GW) or stochastically (BDMC), a subset of the terms 
in a diagrammatic interaction expansion. 
The methods are distinguished by the subsets of diagrams or series terms that are included in the calculation. 
The diagrammatic methods  discussed in this work are 
based on the Feynman diagrammatic technique formulated in terms of self-consistent propagators 
and bare or renormalized interactions \cite{Hedin}. They are formulated at finite temperature but evaluated at low enough temperature that the $T\rightarrow0$ limit can be taken.
\REVISED{Many of the methods were implemented specifically for the present study. The BDMC$_n$
calculations are the first attempt to use high-order skeleton diagrams in materials, to systematically
improve GW.}

\subsubsection{Self-consistent second-order Green's function theory (GF2)}
The fully self-consistent second order Green's function theory (GF2)~\cite{Phillips14,rusakov_gf2_periodic,fractional,GF2_thermo,dahlen04,dahlenmp2} includes all 
second-order skeleton diagrams dressed with the renormalized second-order propagators and bare interactions. GF2  is  formulated as a low-level approximation to the exact 
Luttinger-Ward (LW) functional~\cite{Luttinger1960,Almbladh1999} and therefore is $\Phi$-derivable, thermodynamically consistent, and conserving
\cite{Baym1962, Almbladh1999}. We solve all the non-linear equations self-consistently at non-zero temperature using on an imaginary-time mesh~\cite{Kananenka16,legendre}. 
At each iteration, the self-energy, Green's function, and Fock matrix are updated until convergence is reached, so that the converged solution is reference-independent.

In the weakly correlated regime, GF2 preserves the advantages of the second-order M$\o$ller-Plesset perturbation theory (MP2), while at the same time avoiding the 
divergences appearing in non-self-consistent zero-temperature formulations of finite-order perturbation theories.

We converged the electronic energy to the threshold of $10^{-6} \Ha$. 
The inverse temperature $\beta$ was set at 100 $\Ha^{-1}$ or 200 $\Ha^{-1}$ depending on the geometry. 
The Matsubara freqency grid was generated using splines interpolation \cite{Kananenka16} with between 20,000 and 50,000 points on the Matsubara axis.
The calculations presented here are based on an RHF or UHF reference.
We use GF2 to denote the version that is based on RHF and does not allow for spin symmetry breaking. The acronym
UGF2 is used to denote for a spin unrestricted version based on UHF.

\subsubsection{Diagrammatic methods with renormalized interactions}
The BDMC and SC-GW methods are diagrammatic approximations formulated in terms of renormalized propagators $G$ and renormalized (``screened'') interactions $W$.
They can be written as approximations to the Luttinger-Ward (LW) functional $\Phi$ \cite{Luttinger1960,Almbladh1999}, which implies that
they are thermodynamically consistent and conserving \cite{Baym1962,Almbladh1999}. 
The methods require the self-consistent determination of  propagators $G$, self-energies $\Sigma$, screened interactions $W$, and polarizations $P$. While the expressions for $\Sigma$ and $P$ are different in the individual methods, the Dyson equations
\begin{equation}
G= G_0-G_0\Sigma G \;, \qquad \qquad  W =V+ VPW \, ,
\label{Dyson}
\end{equation}
determine $G$ and $W$, where $G_0$ and $V$ are the bare electronic propagator and interaction.

\subsubsection{Self-consistent GW (SC-GW)}
The self-consistent GW (SC-GW) approximation truncates the skeleton sequence at the lowest-order graph, so that only the first-order contribution in the renormalized interaction is considered and the second-order exchange diagram is neglected.
 We have implemented a deterministic procedure of this approximation closely following Refs.~\cite{Almbladh1999,Stan2006,stan09,Phillips14}:
The Green's function is initialized using the Hartree-Fock (HF) approximation result.
We then construct the polarization $P=GG$ and obtain $W$ from Eq.~(\ref{Dyson}).
After computing the GW self-energy $\Sigma=-GW$, we obtain the updated $G$ by solving Dyson's equation, thus closing the self-consistency loop. 
The method is formulated in a grand canonical ensemble, but the chemical potential $\mu$ used in each step is chosen to preserve the desired electron number. 
After convergence, the  total energy is computed from $\Sigma$ and $G$.

While SC-GW benefits from the conservation of the average particle number, energy, momentum and angular momentum, the size and complexity of $W$ call for 
appropriate controlled simplifications.
Instead of introducing physically motivated approximations that may not respect the conserving properties, we perform systematic linear algebra decompositions and 
truncations on $V$, $W$ and $G$ which vastly reduce the numerical effort \cite{Caruso13}.
We converge our calculations to a relative precision of $10^{-7}$. Convergence is reached at inverse temperature $\beta = 100\,\mathrm{E_{Ha}^{-1}}$, and for about  
$8000$ Matsubara frequencies. A combination of power and uniform meshes, adaptive grids, and spline grids is used for imaginary time and Matsubara data 
\cite{Kananenka16}. The code is based on the ALPS libraries \cite{Gaenko16} and uses integrals generated by LIBINT\cite{Libint2}.

\subsubsection{Treatment of zero-terms in the Hamiltonian}\label{app:zeroorder}
We note an important ambiguity in formulating certain diagrammatic approximations:
Hamiltonian terms that are identically zero because of the Pauli 
principle, and thus have no observable effect on the physical properties of the system, can be arbitrarily added to a Hamiltonian. While these terms will evaluate to zero
in the exact solution, they may evaluate to non-zero values in approximations that do not consider all terms at a given order. The $GW$ approximation is such an approximation, whereas the GF2 approximation does not suffer from this problem.

To illustrate the point, consider an ideal spin-polarized lattice Fermi gas and add a 
contact interaction term
\begin{equation}
 \hat{H}_0 = -\sum_{ij} t_{ij} \hat{\psi}_{j\uparrow}^{\dagger} \hat{\psi}_{i \uparrow} +
 U\sum_{i} \hat{\psi}_{i \uparrow}^{\dagger} \hat{\psi}_{i \uparrow}^{\dagger} 
 \hat{\psi}_{i \uparrow} \hat{\psi}_{i \uparrow}  \, .
\label{polarized_gas}
\end{equation}
The system remains non-interacting, and, correspondingly, in the diagrammatic expansion 
based on the bare interaction vertex $U$, all diagrams of the same order cancel each other 
exactly. However, low-order self-consistent theories like GW may break this cancellation
 by including some, but not all, higher-order contributions in $U$. As a result,
a low-order self-consistent method would produce different answers 
for Eq.~(\ref{polarized_gas}) with $U=0$ and $U\ne 0$.   

\begin{figure}[h!]
\includegraphics[width=0.99\columnwidth]{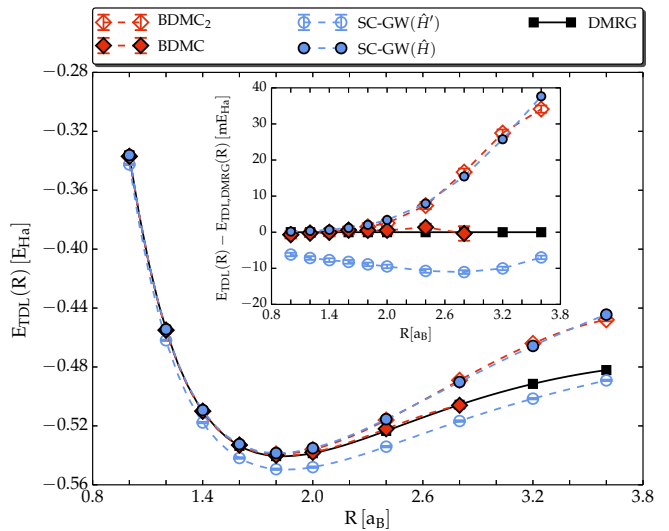}
\caption{Color online: Diagrammatic equation of state in the TDL, at STO-6G level.
Two SC-GW curves underline the ambiguity of formulating the Hamiltonian terms
that are identically zero because of the Pauli principle (see text). The  SC-GW($\hat{H}$)
curve corresponds to the protocol usually used for realistic ab-initio Hamiltonians, while the SC-GW($\hat{H}^\prime$)
curve to the one used in the lattice model Hamiltonian community.
By accounting for higher-order vertex corrections,
the BDMC$_n$ results are observed to converge towards the estimates from other methods.
}
\label{fig:tdl_sto_diag}
\end{figure}

When the interaction Hamiltonian is projected on the orbital basis, similar considerations 
apply to all terms that create or delete two electrons in the same state, meaning that one has 
a choice of keeping or dropping Hamiltonian terms based on matrix elements $v_{abad}$ and $v_{bada}$;
in what follows we will call them ``zero-terms''. In calculations with  realistic Hamiltonians, 
these terms are usually kept, and we will refer
to this choice as Hamiltonian $\hat{H}$. In contrast, the lattice model Hamiltonian community usually omits zero-terms 
explicitly (or nullify the corresponding matrix elements); we will refer to this choice as Hamiltonian
$\hat{H}^\prime$. In an exact solution of the problem, all physical properties of $\hat{H}$ and $\hat{H}^\prime$ are identical.

One question we answer in this work is how the SC-GW results depend on the 
Hamiltonian representation with respect to ``zero-terms''.  The two curves 
labeled as SC-GW($\hat{H}$) and SC-GW($\hat{H}^\prime$) in Fig.~\ref{fig:tdl_sto_diag} correspond to the outcomes 
of the SC-GW method when it is applied to $\hat{H}$ and $\hat{H}^\prime$, respectively. 
The effect of zero-terms is profound. While the SC-GW($\hat{H}$) curve at higher energy 
is more accurate for separations up to the equilibrium distance, it becomes less accurate than the 
SC-GW($\hat{H}^\prime$) curve at $R>2.4 \bohr$, and SC-GW($\hat{H}^\prime$) appears to produce more consistent energy 
differences at the large separation range. 

Given that the two SC-GW answers surround the variational estimate, the difference between 
them can be used as a crude estimate of the accuracy of the SC-GW approximation. This point of view is 
confirmed by our study of vertex corrections. When the second-order vertex corrections are 
accounted for, the result for Hamiltonian $\hat{H}^\prime$ shifts upwards by an amount comparable to
the difference between the SC-GW($\hat{H}$) and SC-GW($\hat{H}^\prime$) curves. The BDMC result starts converging 
to the best variational estimate when higher-order corrections are included, as shown in Fig.~\ref{fig:tdl_sto_diag}.

\subsubsection{Bold diagrammatic Monte Carlo (BDMC)}

We have also developed a stochastic implementation of the $G^{2}W$ formalism that is able to go beyond the lowest-order diagrams.
Within the bold diagrammatic Monte Carlo framework (see e.g. \cite{VanHoucke2010,Kulagin2013,Phonons2016}),
the configuration space of skeleton diagrams for $\Phi$ is sampled stochastically
starting from vertex corrections to SC-GW. 
The method can be applied to any system at non-zero temperature with arbitrary dispersion relation (doped and undoped) 
and with arbitrary shape of the interaction potential \cite{SignBless,VanHoucke2010,Kulagin2013,Phonons2016,Rossi,Rossi2}. 
Both $\Sigma$ and $P$ are computed as sums 
of skeleton graphs, up to order $n$; we denote
these sums as $\Sigma_n$ and $P_n$, and abbreviate the corresponding level of approximation
as BDMC$_n$. The lowest-order contributions to $\Sigma_1$ and $P_1$ are based on products
of the $G$ and $W$ functions mentioned above, and BDMC$_1$ is identical to SC-GW.
To obtain final answers we either perform an extrapolation to the $n \to \infty$ limit, or observe good convergence with increasing
the diagram order. (This was demonstrated for several Coulomb systems in \cite{Phonons2016,DiracL2016}).

In the orbital representation, each interaction line depends on four site/atom
indices $\{(i,j); (k,l)\}$, four orbital indices $\{ (\alpha,\beta);(\gamma,\delta) \}$, and
two spin indices $\{ \sigma; \sigma' \}$ (the Coulomb interaction vertex does not change spin);
to simplify notations we will be also using a composite index $a=(i,\alpha,\sigma )$.
It is worth mentioning that terms with $u= v = 0$, where $u=i-j$ and $v=k-l$ are relative
``distances'' between the orbitals, represent the ``density-density'' part
of the interaction potential, and their contribution is dominating in the final answer.
Accounting for nonzero values of $(u,v)$ in the Dyson equation for $W$ changes the answer at the sub-percent level, and the corresponding contribution quickly
saturates when separation in space
between orbital indices, limited by cutoffs $u^{*}$ and $v^{*}$, is increased.
For hydrogen atoms in the single-orbital case we find that energies per atom obtained with
unrestricted summation over $(u,v)$ and with $u^{*}=v^{*}=2$ coincide at the level of $\sim 10^{-5}$ in relative units even at the smallest values of lattice constant
$R$ considered in this work
(the agreement is better at larger values of $R$).

\section{Tables of results and additional benchmark data}
\label{sec:numeric}

In Tables~\ref{tab:h10_sto6g}, \ref{tab:h10_cbs}, \ref{tab:h10_mrciq}, \ref{tab:tdl_sto},
and \ref{tab:tdl_cbs} we include the
 numerical values of the 
results presented in some of the figures the main text, as well as additional finite-basis and finite-size 
data. Lengths are measured in  Bohr and energies in Hartree. 
Data not included in the appendices will be available online \cite{github2017}.

\begin{table*}[t!]
\resizebox{\textwidth}{!}{%
\begin{tabular}{ccccccccccccccc}
\hline\hline
\begin{tabular}{@{}c@{}} R          \\ $ $       \end{tabular} &
\begin{tabular}{@{}c@{}} AFQMC      \\ $ $       \end{tabular} &
\begin{tabular}{@{}c@{}} DMET[2]    \\ $ $       \end{tabular} &
\begin{tabular}{@{}c@{}} FCI        \\ $ $       \end{tabular} &
\begin{tabular}{@{}c@{}} GF2        \\ $ $       \end{tabular} &
\begin{tabular}{@{}c@{}} SC-GW      \\ $ $       \end{tabular} &
\begin{tabular}{@{}c@{}} SEET(CI/HF)-m  \\ (4,SAO) \end{tabular} &
\begin{tabular}{@{}c@{}} SEET(CI/HF)-m  \\ (6,SAO) \end{tabular} &
\begin{tabular}{@{}c@{}} SEET(CI/GF2)-m \\ (6,SAO) \end{tabular} &
\begin{tabular}{@{}c@{}} RCCSD      \\ $ $       \end{tabular} &
\begin{tabular}{@{}c@{}} RCCSD(T)   \\ $ $       \end{tabular} &
\begin{tabular}{@{}c@{}} UCCSD      \\ $ $       \end{tabular} &
\begin{tabular}{@{}c@{}} UCCSD(T)   \\ $ $       \end{tabular} &
\begin{tabular}{@{}c@{}} RHF        \\ $ $       \end{tabular} &
\begin{tabular}{@{}c@{}} UHF        \\ $ $       \end{tabular} \\
\hline
1.0 & -0.38248(3) & -0.381351 & -0.382439 & -0.381012 & -0.382996 & -0.3817 & -0.3819 & -0.3822 & -0.382387  & -0.382432  & -0.382387  & -0.382432 & -0.375174 & -0.375174  \\
1.2 & -0.47667(3) & -0.475497 & -0.476638 & -0.474570 & -0.476870 & -0.4758 & -0.4760 & -0.4761 & -0.476561  & -0.476626  & -0.476561  & -0.476626 & -0.467805 & -0.467806  \\    
1.4 & -0.52051(5) & -0.519307 & -0.520509 & -0.517622 & -0.520234 & -0.5196 & -0.5199 & -0.5200 & -0.520401  & -0.520492  & -0.520401  & -0.520492 & -0.509862 & -0.509862  \\     
1.6 & -0.53819(6) & -0.537170 & -0.538436 & -0.534488 & -0.537403 & -0.5374 & -0.5378 & -0.5379 & -0.538292  & -0.538414  & -0.538292  & -0.538414 & -0.525628 & -0.525628  \\     
1.8 & -0.54218(6) & -0.541097 & -0.542439 & -0.537111 & -0.540312 & -0.5413 & -0.5418 & -0.5418 & -0.542254  & -0.542418  & -0.541742  & -0.542175 & -0.527014 & -0.527745  \\     
2.0 & -0.53874(5) & -0.537527 & -0.538963 & -0.531852 & -0.535304 & -0.5376 & -0.5383 & -0.5383 & -0.538745  & -0.538966  & -0.537504  & -0.538049 & -0.520347 & -0.523137  \\     
2.4 & -0.52271(8) & -0.521125 & -0.522794 & -0.510567 & -0.514336 & -0.5212 & -0.5221 & -0.5221 & -0.522681  & -0.523088  & -0.520417  & -0.520753 & -0.495606 & -0.506571  \\     
2.8 & -0.50507(5) & -0.503220 & -0.505024 & -0.485323 & -0.488955 & -0.5034 & -0.5044 & -0.5045 & -0.506729  & -0.507486  & -0.502820  & -0.502959 & -0.465843 & -0.491183  \\    
3.2 & -0.49114(2) & -0.489450 & -0.491038 & -0.461886 & -0.464687 & -0.4897 & -0.4906 & -0.4908 & -0.590076  & -0.592683  & -0.489482  & -0.489521 & -0.436679 & -0.481323  \\     
3.6 & -0.48200(2) & -0.480818 & -0.481870 & -0.442549 & -0.443705 & -0.4808 & -0.4816 & -0.4819 & -0.664459  & -0.668234  & -0.480951  & -0.480959 & -0.410493 & -0.476035  \\
\hline\hline
\end{tabular}
}
\caption{Potential energy curve of \chem{H_{10}} with the minimal (STO-6G) basis. 
DMET[5], MRCI and MRCI+Q energies coincide with FCI to within $10^{-6}$.} \label{tab:h10_sto6g}
\end{table*}
\begin{table*}[t!]
\resizebox{\textwidth}{!}{%
\begin{tabular}{ccccccccc}
\hline\hline
R   & AFQMC        & DMET[2]      & LR-DMC(AGP)  &  LR-DMC(LDA)  & MRCI+Q       & MRCI+Q+F12   & PC-NEVPT2   & SC-NEVPT2   \\
\hline                                                                                                                  
1.0 & -0.44284(24) & N/A          & -0.442430(7) &  -0.442093(9) & -0.44324(32) & -0.44301(31) & N/A         & N/A         \\
1.2 & -0.51489(14) & N/A          & -0.514432(6) &  -0.514418(7) & -0.51529(15) & -0.51506(12) & N/A         & N/A         \\
1.4 & -0.54914(11) & -0.54905(28) & -0.548742(6) &  -0.548748(7) & -0.54926(15) & -0.54917(12) & -0.54593(3) & -0.54569(3) \\
1.6 & -0.56315(12) & -0.56258(30) & -0.562878(6) &  -0.562900(7) & -0.56324(8)  & -0.56321(7)  & -0.56011(1) & -0.55991(1) \\
1.8 & -0.56644(4)  & -0.56567(22) & -0.566234(6) &  -0.566255(7) & -0.56655(5)  & -0.56654(5)  & -0.56350(5) & -0.56333(6) \\
2.0 & -0.56396(6)  & -0.56321(12) & -0.563831(5) &  -0.563852(7) & -0.56411(3)  & -0.56410(3)  & -0.56113(7) & -0.56099(8) \\
2.4 & -0.55164(4)  & -0.55124(2)  & -0.551679(6) &  -0.551699(8) & -0.55189(1)  & -0.55187(1)  & -0.54911(8) & -0.54902(9) \\
2.8 & -0.53755(8)  & -0.53726(2)  & -0.537423(6) &  -0.537424(9) & -0.53754(3)  & -0.53753(3)  & -0.53510(9) & -0.53504(9) \\
3.2 & -0.52499(7)  & -0.52512(6)  & -0.524987(7) &  -0.524974(9) & -0.52499(6)  & -0.52500(5)  & -0.52302(9) & -0.52299(9) \\
3.6 & -0.51568(9)  & -0.51612(9)  & -0.515571(7) & -0.515582(11) & -0.51549(8)  & -0.51553(5)  & -0.51405(8) & -0.51403(8) \\
\hline\hline
\\
\hline\hline
R   & SBDMRG       & RCCSD        & RCCSD(T)     & UCCSD        & UCCSD(T)     & UHF           & VMC(AGP)     & VMC(LDA)     \\
\hline                                                                                                                       
1.0 & -0.44209(21) & -0.44207(26) & -0.44268(21) & -0.44207(26) & -0.44268(21) & -4.15363(16)  & -0.441499(9) & -0.44061(1)  \\
1.2 & -0.51399(15) & -0.51432(19) & -0.51490(19) & -0.51432(19) & -0.51490(19) & -4.88368(5)   & -0.513646(9) & -0.513225(8) \\
1.4 & -0.54859(15) & -0.54846(14) & -0.54907(14) & -0.54846(14) & -0.54907(14) & -5.22845(4)   & -0.548046(7) & -0.547650(8) \\
1.6 & -0.56311(14) & -0.56240(10) & -0.56311(9)  & -0.56240(10) & -0.56311(9)  & -5.36820(3)   & -0.562237(5) & -0.561823(7) \\
1.8 & -0.56666(14) & -0.56565(4)  & -0.56638(4)  & -0.56533(4)  & -0.56623(4)  & -5.40150(2)   & -0.565634(5) & -0.565201(6) \\
2.0 & -0.56428(14) & -0.56311(1)  & -0.56391(1)  & -0.56239(1)  & -0.56341(1)  & -5.38220(2)   & -0.563249(4) & -0.562776(6) \\
2.4 & -0.55196(14) & -0.55071(2)  & -0.55167(3)  & -0.54953(2)  & -0.55037(2)  & -5.281261(6)  & -0.551106(4) & -0.550575(5) \\
2.8 & -0.53765(14) & -0.53632(5)  & -0.53749(6)  & -0.53519(5)  & -0.53577(6)  & -5.174240(5)  & -0.536832(3) & -0.536201(5) \\
3.2 & -0.52511(14) & -0.52419(8)  & -0.52574(10) & -0.52303(8)  & -0.52340(8)  & -5.096702(6)  & -0.524373(4) & -0.523557(5) \\
3.6 & -0.51568(14) & N/A          & N/A          & -0.51407(10) & -0.51428(10) & -5.050387(7)  & -0.514968(3) & -0.513869(6) \\
\hline\hline 
\end{tabular}
}
\caption{Potential energy curve of \chem{H_{10}} extrapolated to the CBS limit. 
The MRCI+Q value at $R=1.0$ uses the AFQMC energy with a correction 
estimated from the difference between  MRCI+Q  and AFQMC energies at $R=1.2$.
RCC breaks down at $R=3.6$ and, for large basis sets and the shortest bondlengths, DMET[2], PC-NEVPT2 and SC-NEVPT2
are unconvergent due to linear dependency issues.
} \label{tab:h10_cbs}
\end{table*}

\begin{table}[h!]
\begin{tabular}{ccccc}
\hline\hline
R   & cc-pVDZ   & cc-pVTZ   & cc-pVQZ   & cc-pV5Z   \\
\hline
1.0 & -0.421954 & N/A	    & N/A	& N/A       \\
1.2 & -0.502812 & -0.513118 & N/A	& N/A       \\
1.4 & -0.540966 & -0.547756 & -0.548548 & N/A       \\
1.6 & -0.557110 & -0.561985 & -0.562710 & -0.562932 \\
1.8 & -0.561486 & -0.565330 & -0.566068 & -0.566277 \\
2.0 & -0.559552 & -0.562892 & -0.563645 & -0.563852 \\
2.4 & -0.547661 & -0.550663 & -0.551426 & -0.551651 \\
2.8 & -0.533470 & -0.536307 & -0.537099 & -0.537349 \\
3.2 & -0.521270 & -0.523817 & -0.524586 & -0.524858 \\
3.6 & -0.512397 & -0.514463 & -0.515150 & -0.515425 \\
 \hline\hline
\end{tabular}
\caption{Potential energy curve of \chem{H_{10}} at cc-pVxZ, from the MRCI+Q method.
For large basis sets and the shortest bondlengths, calculations
are unconvergent due to linear dependency issues.
Additional calculations were performed with F12 corrections.
} \label{tab:h10_mrciq}
\end{table}

\

\begin{table*}[t!]
\resizebox{\textwidth}{!}{%
\begin{tabular}{ccccccccccccccc}
\hline\hline
\begin{tabular}{@{}c@{}} R              \\ $ $       \end{tabular} &
\begin{tabular}{@{}c@{}} AFQMC          \\ $ $       \end{tabular} &
\begin{tabular}{@{}c@{}} BDMC           \\ $ $       \end{tabular} &
\begin{tabular}{@{}c@{}} DMET[5]        \\ $ $       \end{tabular} &
\begin{tabular}{@{}c@{}} DMET[$\infty$] \\ $ $       \end{tabular} &
\begin{tabular}{@{}c@{}} DMRG           \\ $ $       \end{tabular} &
\begin{tabular}{@{}c@{}} SC-GW          \\ $ $       \end{tabular} &
\begin{tabular}{@{}c@{}} SEET(CI/GF2)-m \\ (6,SAO)   \end{tabular} &
\begin{tabular}{@{}c@{}} SEET(CI/HF)-m  \\ (6,SAO)   \end{tabular} &
\begin{tabular}{@{}c@{}} RCCSD          \\ $ $       \end{tabular} &
\begin{tabular}{@{}c@{}} RCCSD(T)       \\ $ $       \end{tabular} &
\begin{tabular}{@{}c@{}} UCCSD          \\ $ $       \end{tabular} &
\begin{tabular}{@{}c@{}} UCCSD(T)       \\ $ $       \end{tabular} &
\begin{tabular}{@{}c@{}} UHF            \\ $ $       \end{tabular} \\
\hline  
1.0 & -0.33631(5) & -0.337(1) & -0.33730 & -0.33621 & -0.33631 & -0.33614 & -0.3353 & -0.3354 & -0.33596 & -0.33625 & -0.33596 & -0.33625 & -0.32742 \\
1.2 & -0.45466(5) & -0.455(1) & -0.45569 & -0.45470 & -0.45474 & -0.45439 & -0.4537 & -0.4536 & -0.45428 & -0.45464 & -0.45385 & -0.45437 & -0.44504 \\
1.4 & -0.50999(4) & -0.510(1) & -0.51081 & -0.50989 & -0.50996 & -0.50930 & -0.5088 & -0.5088 & -0.50937 & -0.50981 & -0.50906 & -0.50939 & -0.49958 \\
1.6 & -0.53356(5) & -0.533(1) & -0.53428 & -0.53342 & -0.53356 & -0.53238 & -0.5323 & -0.5323 & -0.53286 & -0.53337 & -0.53278 & -0.53310 & -0.52218 \\
1.8 & -0.54048(4) & -0.540(1) & -0.54104 & -0.54034 & -0.54049 & -0.53844 & -0.5391 & -0.5390 & -0.53967 & -0.54026 & -0.53962 & -0.54001 & -0.52806 \\
2.0 & -0.53846(7) & -0.538(1) & -0.53888 & -0.53840 & -0.53850 & -0.53512 & -0.5369 & -0.5371 & -0.53760 & -0.53833 & -0.53740 & -0.53780 & -0.52517 \\
2.4 & -0.52326(4) & -0.522(1) & -0.52350 & -0.52335 & -0.52336 & -0.51544 & -0.5219 & -0.5218 & N/A      & N/A      & -0.52182 & -0.52211 & -0.50920 \\
2.8 & -0.50556(5) & -0.506(2) & -0.50578 & -0.50564 & -0.50564 & -0.49018 & -0.5050 & -0.5043 & N/A      & N/A      & -0.50406 & -0.50420 & -0.49305 \\
3.2 & -0.49145(4) & N/A       & -0.49166 & -0.49145 & -0.49146 & -0.46570 & -0.4920 & -0.4904 & N/A      & N/A      & -0.49026 & -0.49030 & -0.48238 \\
3.6 & -0.48210(5) & N/A       & -0.48240 & -0.48210 & -0.48211 & -0.44446 & -0.4836 & -0.4816 & N/A      & N/A      & -0.48137 & -0.48138 & -0.47657 \\
\hline\hline
\end{tabular}
}
\caption{Equation of state of the hydrogen chain at the TDL, computed with the minimal basis (STO-6G).
} \label{tab:tdl_sto}
\end{table*}

\begin{table*}[t!]
\resizebox{\textwidth}{!}{%
\begin{tabular}{cccccccccccc}
\hline\hline
R   & AFQMC        & AFQMC+$\Delta_{\mathrm{DMRG}}$ & DMET[2]      & LR-DMC(LDA)  & SC-NEVPT2     & RCCSD        & RCCSD(T)     &  UCCSD       & RHF          &  UHF         & VMC(LDA)     \\
\hline
1.4 & -0.54044(35) & -0.54022(35)       & -0.53858(27) & -0.53971(6)  & -0.53674(7)   & -0.53895(19) & -0.54009(19) & -0.53897(22) & -0.51133(3)  & -0.51234(3)  & -0.53820(1)  \\
1.6 & -0.55971(36) & -0.55959(37)       & -0.55730(34) & -0.55912(12) & -0.55608(11)  & -0.55821(13) & -0.55942(14) & -0.55778(15) & -0.53097(2)  & -0.53256(2)  & -0.55767(7)  \\
1.8 & -0.56569(30) & -0.56562(31)       & -0.56312(26) & -0.56515(3)  & -0.56220(4)   & -0.56401(6)  & -0.56529(5)  & -0.56354(7)  & -0.53685(2)  & -0.53925(1)  & -0.56346(1)  \\
2.0 & -0.56444(34) & -0.56451(34)       & -0.56213(14) & -0.56397(4)  & -0.56117(8)   & -0.56272(1)  & -0.56406(3)  & -0.56238(2)  & -0.53535(1)  & -0.53894(1)  & -0.56216(1)  \\
2.4 & -0.55313(31) & -0.55291(31)       & -0.55135(1)  & -0.55268(4)  & -0.54999(9)   & -0.55107(1)  & -0.55257(3)  & -0.55105(2)  & -0.52235(1)  & -0.52978(1)  & -0.55004(1)  \\
2.8 & -0.53886(29) & -0.53870(30)       & -0.53768(1)  & -0.53848(5)  & -0.53594(10)  & N/A          & N/A	    & -0.53666(7)  & -0.50495(1)  & -0.51883(1)  & -0.53482(1)  \\
3.2 & -0.52557(23) & -0.52549(23)       & -0.52566(5)  & -0.52585(5)  & -0.52362(10)  & N/A          & N/A	    & -0.52403(12) & -0.48754(1)  & -0.51058(1)  & -0.52250(1)  \\
3.6 & -0.51611(22) & -0.51594(22)       & -0.51685(8)  & -0.51608(9)  & -0.51443(9)   & N/A          & N/A	    & -0.51462(14) & -0.47157(4)  & -0.50556(1)  & -0.51273(1)  \\
\hline\hline
\end{tabular}
}
\caption{Final equation of state for the hydrogen chain at the thermodynamic limit. \REV{LR-DMC results for $R=1.0$ and $1.2$ are $-0.4198(3)$ and $-0.4987(6)$, respectively.}
} \label{tab:tdl_cbs}
\end{table*}

\begin{figure*}[t!]
\centering
\includegraphics[width=0.65\textwidth]{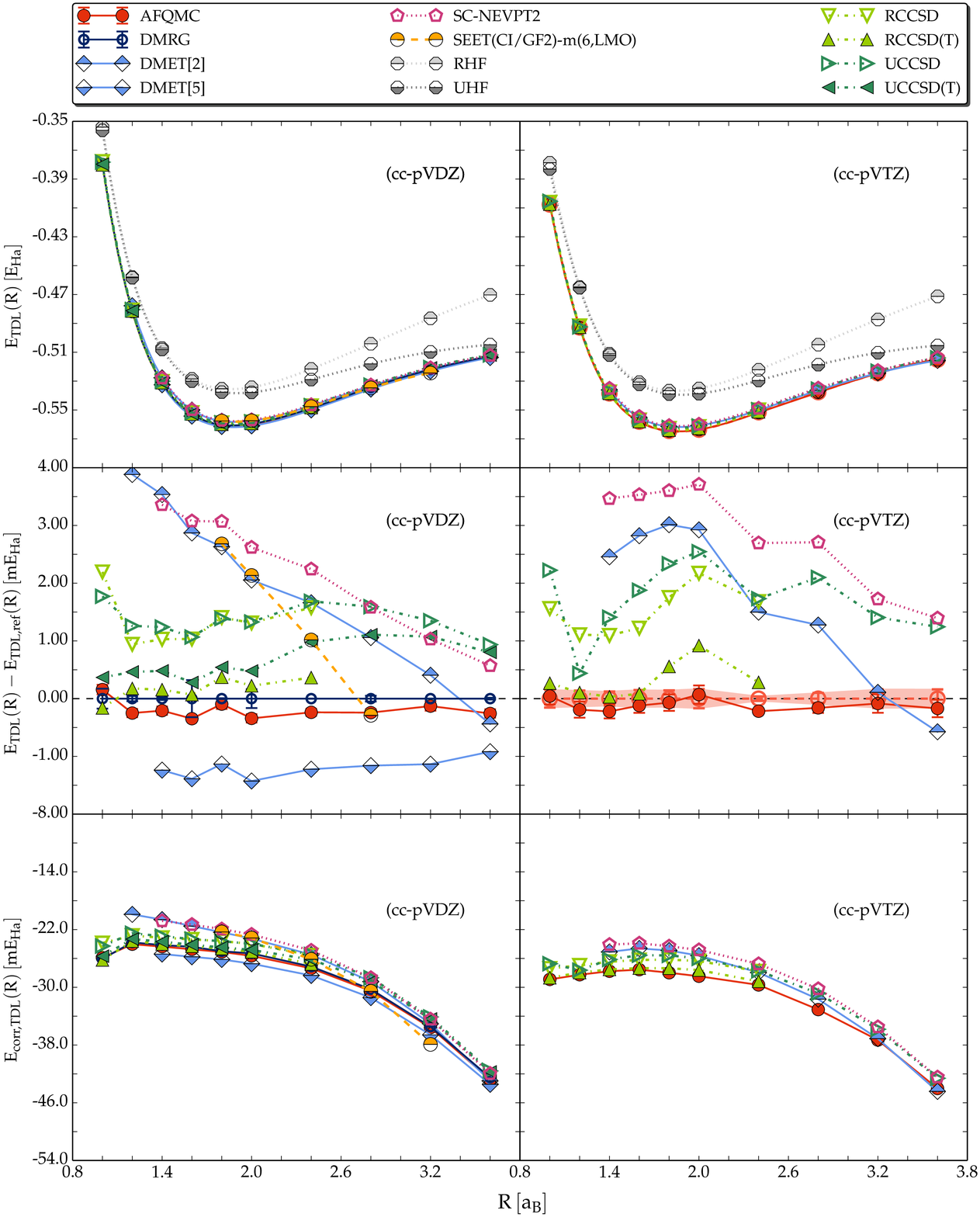}
\caption{
Top: EOS in the thermodynamic limit computed with finite basis sets (cc-pVDZ  and cc-pVTZ). 
Middle: detailed comparison using DMRG and AFQMC as reference. 
Error bars on the DMRG data indicate estimates of the TDL extrapolation uncertainties 
based on the results at the STO-6G level in Fig.~\ref{fig:tdl_sto_extra2}.
AFQMC+$\Delta$DMRG$_{\rm DZ}$ is shown as reference for TZ (empty red circle),
where the correction is obtained from the energy difference between DMRG and AFQMC 
at DZ. Bottom: Correlation energy per particle in the TDL, at cc-pVDZ (left) and cc-pVTZ (right) level.
} \label{fig:tdl_xz}
\end{figure*}

\begin{figure}[h!]
\centering
\includegraphics[width=0.99\columnwidth]{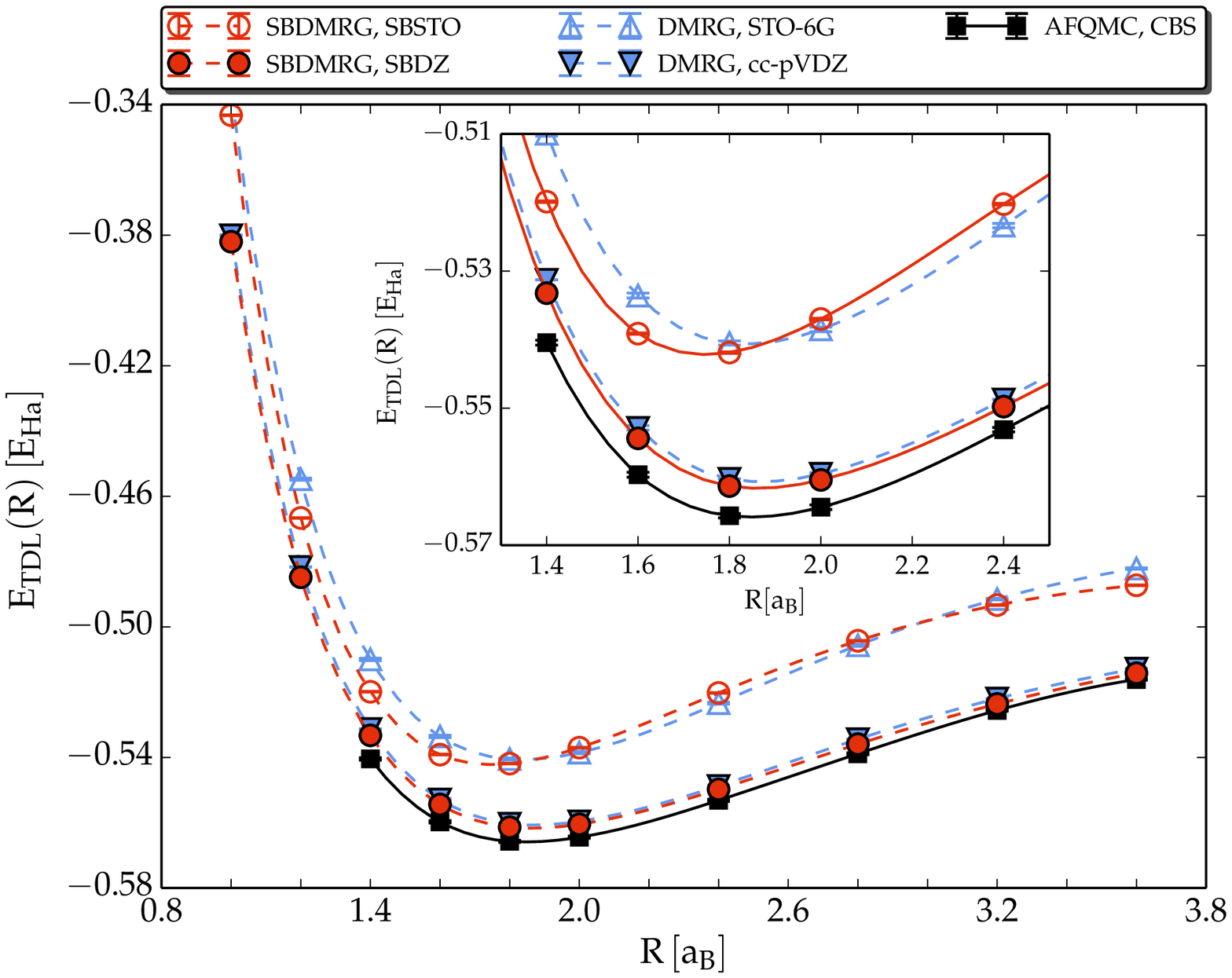}
\caption{
Equation of state in the thermodynamic limit with select finite basis sets, by DMRG and SBDMRG. 
AFQMC results at the CBS limit are also shown for reference.
} \label{fig:ggdmrg_tdl}
\end{figure}

A variety of finite-size and/or finite-basis-set data are available. 
Figure~\ref{fig:tdl_xz} shows the equation of state (EOS) extrapolated to the TDL at cc-pVDZ and cc-pVTZ level.
At cc-pVDZ level, DMRG provides a highly accurate EoS, with equilibrium bondlength and energy
$R_{eq} = 1.880(2) \bohr$, $E_{0} = -0.5608(2) \Ha$. The corresponding correlation energies, using RHF energies 
as reference, are shown in the middle panel of Fig.~\ref{fig:tdl_xz}.
Figure~\ref{fig:ggdmrg_tdl}
shows the EOS extrapolated to the TDL using SBDMRG and DMRG for STO-6G  and cc-pVDZ
level basis sets.

\section{Additional details on reaching the complete basis set and thermodynamic limits}
\label{sec:H10cbsappendix}

\subsection{Extrapolation to the CBS limit}

\begin{figure*} 
\centering
\includegraphics[width=0.9\textwidth]{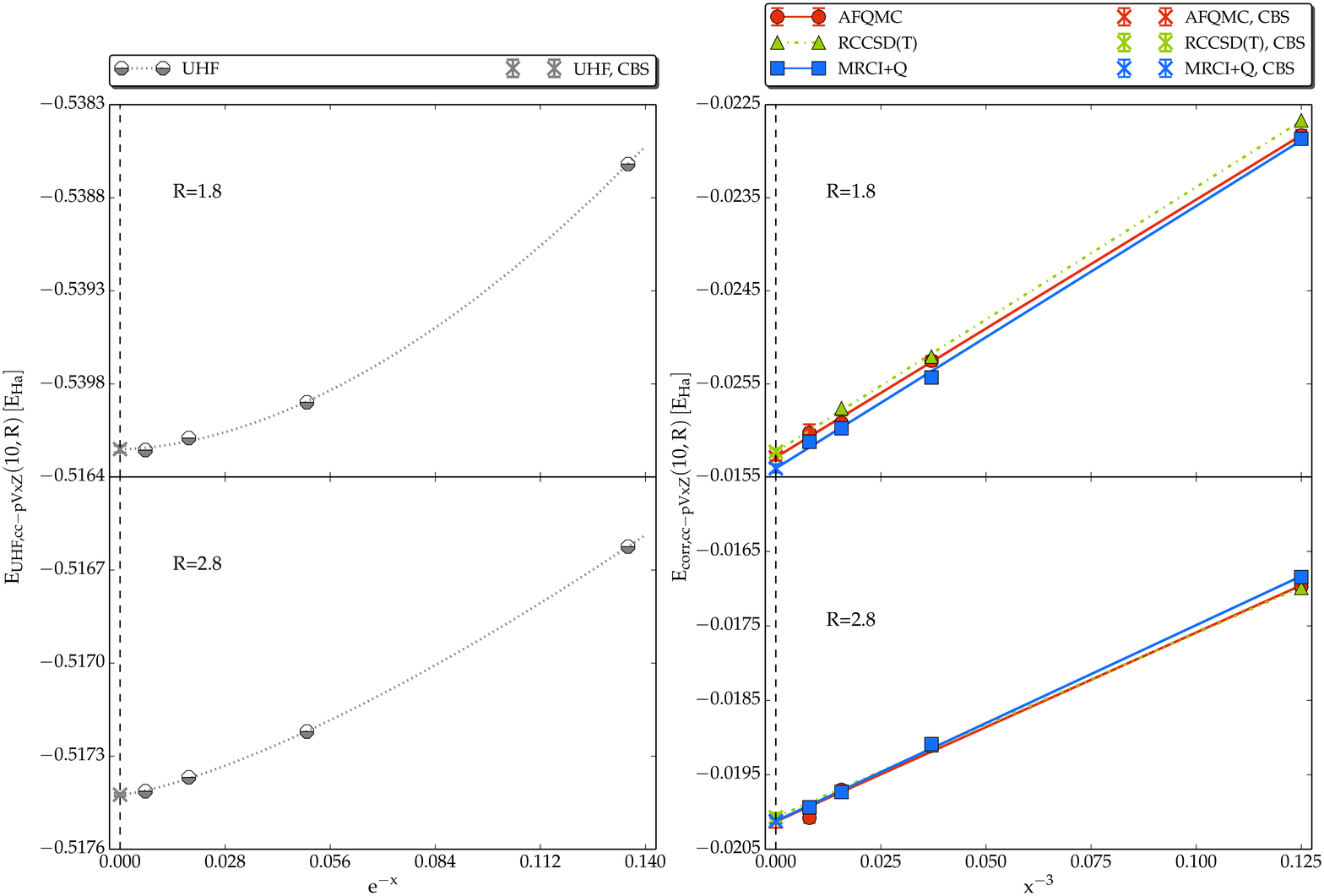}
\caption{
Illustration of the 
extrapolation to the CBS limit. 
Results are shown for \chem{H_{10}} at bondlengths, $R=1.8$ and $2.8 \bohr$.
The unrestricted Hartree-Fock energy is fitted to an exponential function of the index x=2,3,4,5 of the cc-pVxZ basis,
the correlation energy to the power law $\alpha + \beta \, x^{-3}$.
} \label{fig:h10_to_cbs}
\end{figure*}

Extrapolations of the UHF and UHF-based correlation energy to the CBS limit are illustrated in Fig.~\ref{fig:h10_to_cbs}, for 
the representative bondlengths $R=1.8$, $2.8 \bohr$. Using RHF as references gives indistinguishable results.

\begin{figure}[h!]
\centering
\includegraphics[width=0.85\columnwidth]{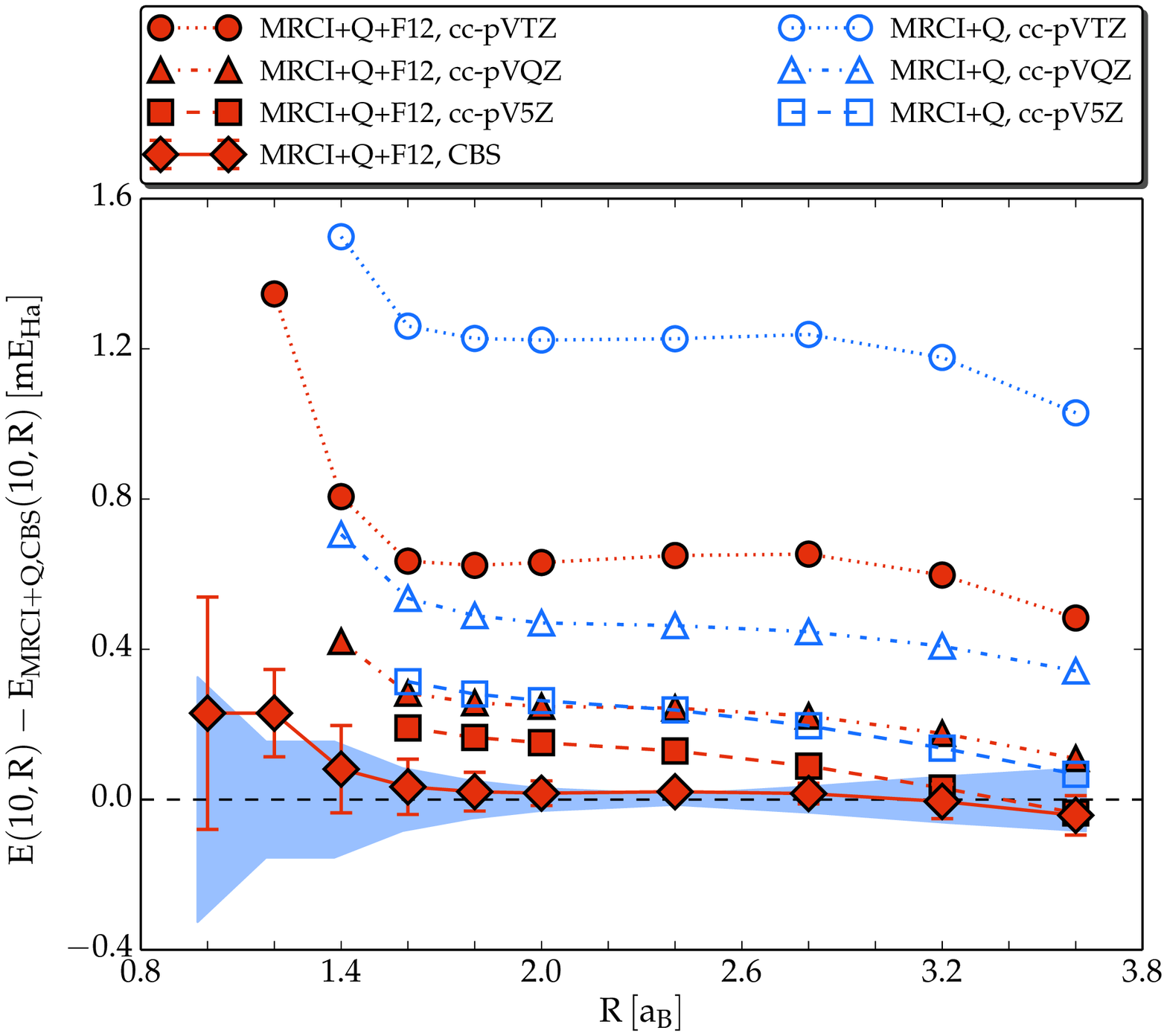}
\caption{
Effect of the F12 correction on MRCI+Q data.
A cc-pVxZ+F12 calculation yields an energy of cc-pV(x+1)Z quality.
Extrapolations to the CBS agree with each other within uncertainties from the fitting procedure.
} \label{fig:h10_f12}
\end{figure}

Figure~\ref{fig:h10_f12} shows the effect of the F12 correction on MRCI+Q energies.
As illustrated in the main text, 
extrapolations to the CBS limit obtained with and without F12 correction agree with each other to within the fitting uncertainties,
confirming the robustness of the extrapolation procedure.

\subsection{Extrapolation to the TDL}
\label{sec:extrapolation_technique}

\begin{figure}[h!]
\centering
\includegraphics[width=0.9\columnwidth]{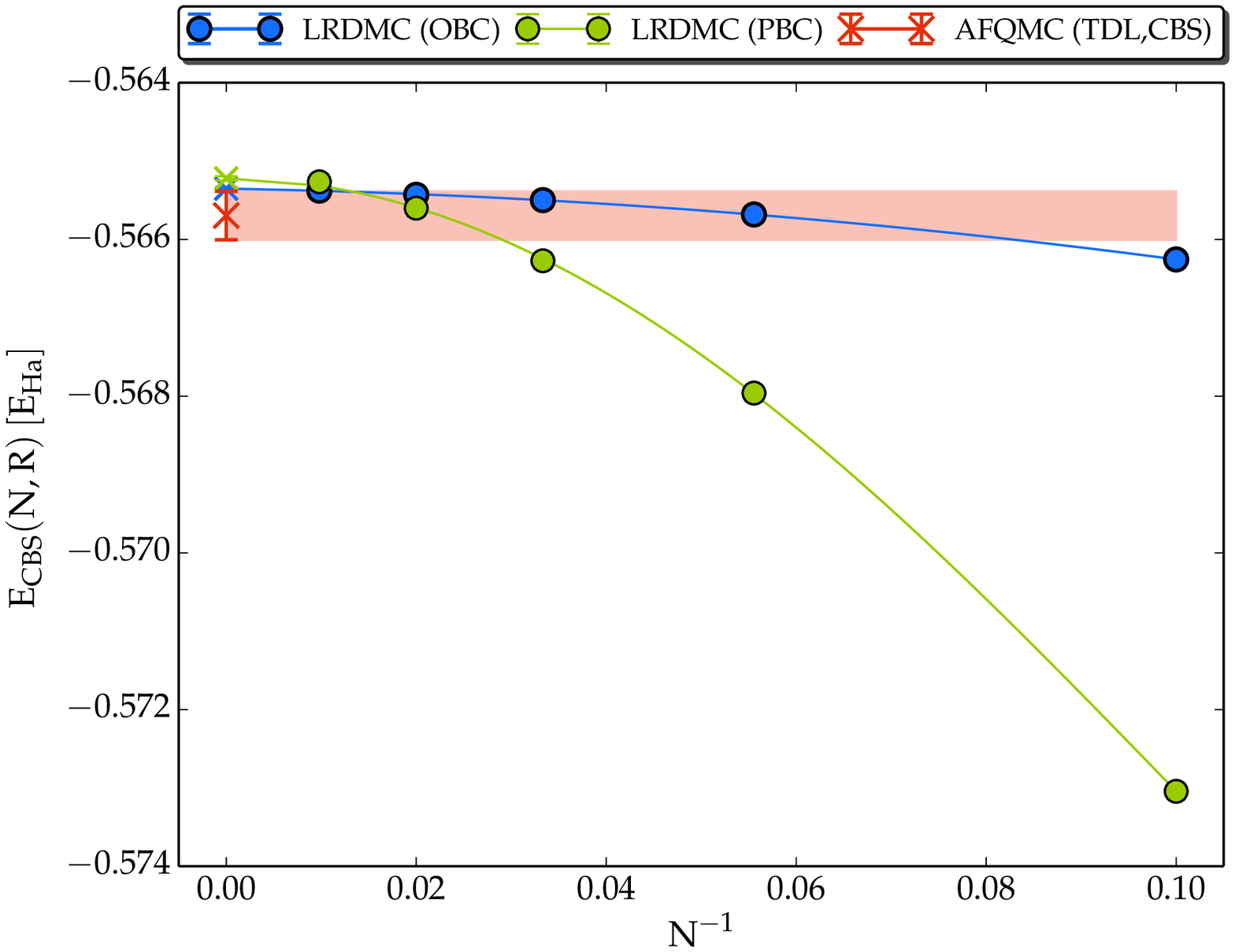}
\caption{
Extrapolation to the thermodynamic limit  using periodic (PBC) vs.~open (OBC) boundary conditions.
Results are shown from LR-DMC for $R=1.8 \bohr$. Both sets of calculations converge to the same 
result for $N\to \infty$, but convergence is more rapid with OBC. The final result for $R=1.8 \bohr$ from AFQMC 
(from Fig.~\ref{tdl_cbs}) is also indicated for reference.
} \label{fig_dmc_bc}
\end{figure}

As mentioned, we find that the use of finite clusters (chains) tends to give better convergence to the 
TDL than with periodic boundary conditions, except for very short bondlengths. This is illustrated in Fig.~\ref{fig_dmc_bc}. 
The faster convergence 
with OBC than PBC is somewhat surprising and counter to commonly held belief.
The quasi-one-dimensional nature of the hydrogen chain is likely an important factor.

\begin{figure*}[t!]
\centering
\includegraphics[width=0.75\textwidth]{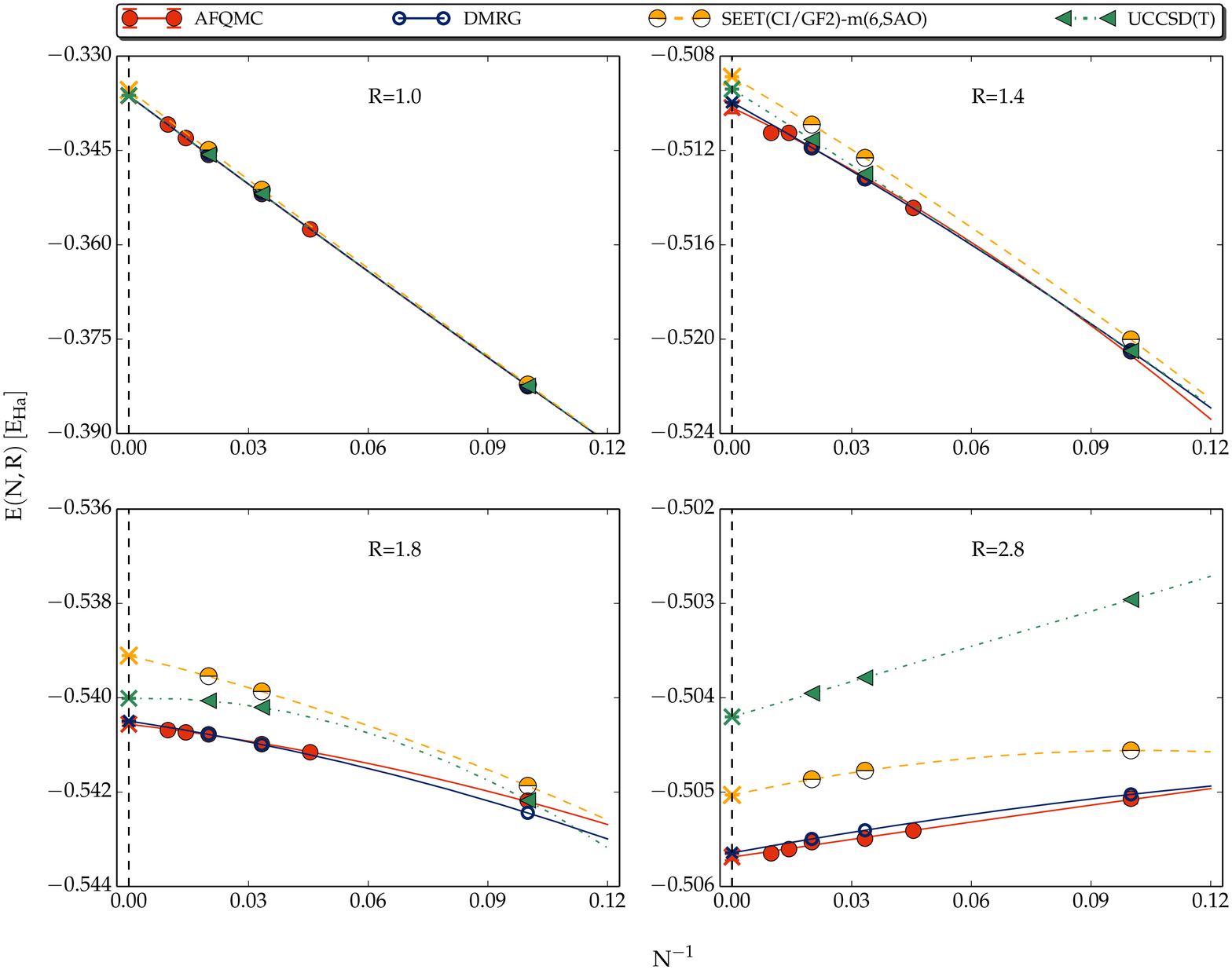}
\caption{
Extrapolation of the EOS to the TDL limit, in the minimal STO-6G basis.
The energy per particle is fitted to a second-order polynomial $A_0 + A_1 N^{-1} + A_2 N^{-2}$ in $N^{-1}$,
reflecting the presence of a bulk and a boundary contribution.
} \label{fig:tdl_sto_extra}
\end{figure*}

Extrapolations to the TDL are illustrated in the minimal basis in Fig.~\ref{fig:tdl_sto_extra} for the bondlengths 
$R=1.0$, $1.4$, $1.8$, $2.8$.
The importance of the $A_2(R) N^{-2}$ correction is evident in capturing the size effects.
Note that the finite-size effects are larger at shorter bondlengths (ranging from roughly $50\,\mHa$ for 
$R=1.0 \bohr$ to less than $1\,\mHa$ at large separation as the chain turns into 
a collection of uncoupled \chem{H} atoms).
This is expected from the 
nature of the long-range Coulomb interaction.

\begin{figure*}[t!]
\centering
\includegraphics[width=0.75\textwidth]{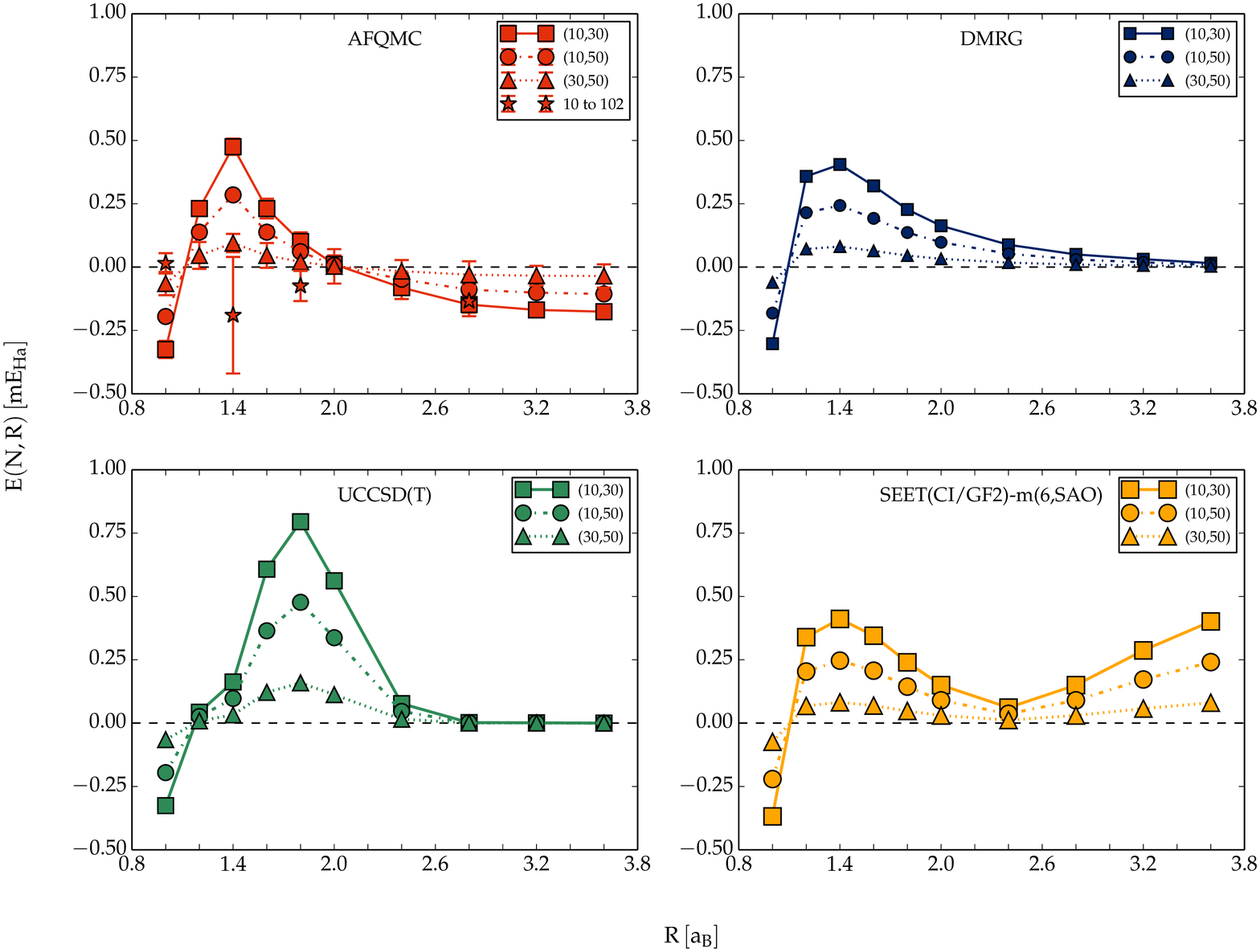}
\caption{
Illustration of the extrapolation to TDL. 
Residual errors of fits from different pairs $(N_1,N_2)$ of finite sizes are shown versus the 
reference extrapolation from $N=10$, $30$, $50$. Data are for the  minimal STO-6G basis.
Large system sizes are investigated in some cases to verify convergence.
} \label{fig:tdl_sto_extra2}
\end{figure*}

In Fig.~\ref{fig:tdl_sto_extra2}, we examine the robustness of the extrapolation.
Results from the simple subtraction trick separating surface and bulk [i.e. $k=1$ in Eq.~(\ref{eq:tdl_xtrap})]
are compared with the reference extrapolation using $E_{N_1,N_2,N_3}(R)$.
Extrapolations agree with each other to well within 1 $\mHa$, and approach 
$E_{N_1,N_2,N_3}(R)$ as $N_1$, $N_2$ are increased.
These results suggest  that extrapolations 
$E_{N_1,N_2,N_3}(R)$ to the TDL have a resolution of the order of $0.1 \mHa$ per particle.
For most methodologies, therefore, the uncertainty on the TDL extrapolation is one order of 
magnitude smaller than the bias due to the underlying approximations. It is also well within the 
uncertainty bound in our final best estimate of the EOS.

Our DMET strategy for calculations that directly access the TDL is
somewhat different. We start from a RHF calculation on a large system
(H$_{150}$, H$_{300}$, \ldots). A single fragment around the central
H-atom is constructed, using $x-1$ neighbors around it. The energy
contribution due to the central H-atom is taken as the energy per
atom, while the chemical potential is adjusted in the central H-atom
such that its particle number contribution is 1.

In order to more efficiently access large impurity sizes, we truncate
the embedding space such that the bath orbitals with a small norm ($<
0.01$) are excluded from the high level calculation. Our DMET results
are reported using DMRG as a solver and a bond dimension of $D =
1000$. The error from the DMRG solver is less than
$1$~$\mu$E$_{\textrm{Ha}}$ in the energy per atom. To converge to the
TDL it is necessary to converge both the full system as well as the
fragment size. We find that it is necessary for the full system to be
very large to converge to the TDL at short bond lengths. We carry out
calculations on systems of increasing size (up to H$_{1950}$ at $R =
1.0$~a$_{\textrm{B}}$) using a fixed fragment size [22], until the
change in the energy per atom is smaller than
$0.01$~mE$_{\mathrm{Ha}}$. Using this suitably defined full system
size \footnote{For $R=1.0$ and $1.2$~a$_{\textrm{B}}$ we performed
  fragment size extrapolations on H$_{1200}$, which is not converged
  with respect to the full system size. Our extrapolated result
  includes a correction due to the full system size determined from
  calculations on larger systems with a fragment of size [22].}, we
perform calculations on larger and larger fragments (see Fig. 17,
panel b) and perform a quadratic extrapolation with the inverse of
fragment size. The fragment size extrapolation using the largest 4 or
5 fragment sizes yields the same limit to better than
$0.02$~mE$_{\mathrm{Ha}}$.

\begin{figure}[!htb]
  \includegraphics[width=0.85\columnwidth]{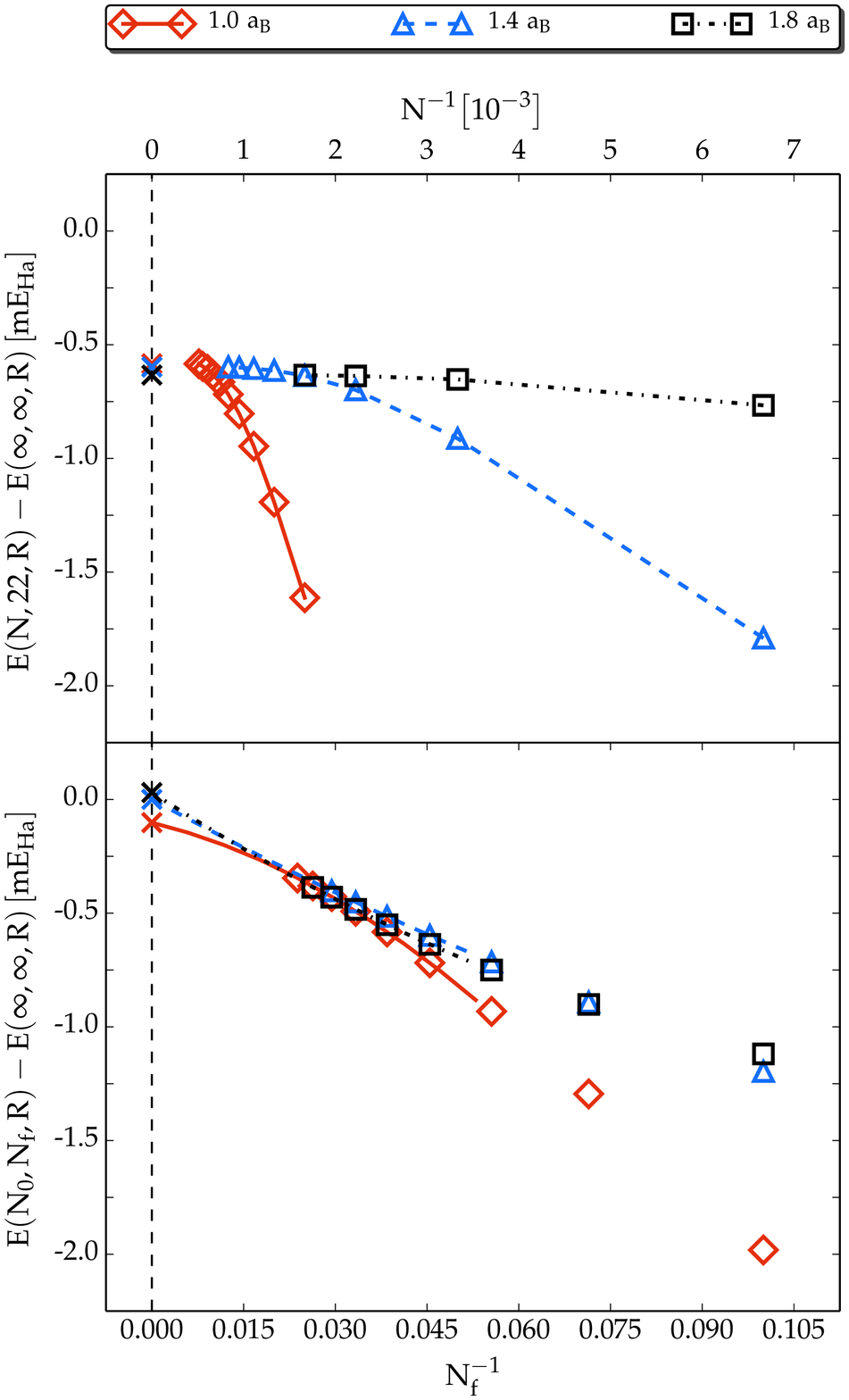}
  \caption{Top: difference with respect to extrapolated energies $E_{\tdl}(R) \equiv E(\infty,\infty,R)$ in
    STO-6G DMET calculations of increasing full system size $N$ (using a
    fixed fragment of size [22]) for three different geometries. (b)
    Difference with respect to extrapolated energies $E_{\tdl}(R) \equiv E(\infty,\infty,R)$ in STO-6G DMET
    calculations of increasing fragment size $N_f$ (using a fixed full
    system size $N_0$: $\chem{H_{1200}}$ for $1.0$ and $1.4 \bohr$ and $\chem{H_{450}}$ for
    $1.8 \bohr$). The difference between the asymptotic value $E(N_0,\infty,R)$ and the final extrapolation
    $E(\infty,\infty,R)$ is estimated as $E(\infty,\infty,R)-E(N_0,\infty,R) \simeq E(\infty,22,R)-E(N_0,22,R)$
    \label{fig-dmet}}
\end{figure}

\bibliographystyle{ieeetr}

\end{document}